\documentclass[twocolumn,aps,prb,superscriptaddress,floatfix,amsmath,amssymb,preprintnumbers]{revtex4-1}
\usepackage{graphicx}
\usepackage{dcolumn}
\usepackage{bm}
\usepackage{float}
\usepackage{url}
\usepackage{amssymb}
\usepackage{calc}
\usepackage{latexsym}
\usepackage{epsf}
\usepackage{epsfig}
\usepackage{notoccite}
\usepackage{euscript}
\usepackage{hyperref}
\usepackage{xcolor}
\usepackage{sidecap}

\begin{document}

\title{Probing Anomalous Inelastic Scattering in Spin-ice Ho$_2$Ti$_2$O$_7$ through \\ Resonant Raman Spectroscopy}

\author{Naween Anand}
\affiliation{National High Magnetic Field Laboratory, FSU, Tallahassee, FL 32310, USA}

\author{L. J. van de Burgt}
\affiliation{Department of Chemistry, FSU, Tallahassee, FL 32310, USA}

\author{Q. Huang}
\affiliation{University of Tennessee, Knoxville, TN 37996, USA}

\author{Jade Holleman}
\affiliation{National High Magnetic Field Laboratory, FSU, Tallahassee, FL 32310, USA}
\affiliation{Department of Physics, FSU, Tallahassee, FL 32310, USA}

\author{Haidong Zhou}
\affiliation{University of Tennessee, Knoxville, TN 37996, USA}

\author{Stephen A McGill}
\affiliation{National High Magnetic Field Laboratory, FSU, Tallahassee, FL 32310, USA}

\author{Christianne Beekman}
\affiliation{National High Magnetic Field Laboratory, FSU, Tallahassee, FL 32310, USA}
\affiliation{Department of Physics, FSU, Tallahassee, FL 32310, USA}

\date{\today}

\begin{abstract}
We report on a study of the inelastic scattering properties of (001) and (111) Ho$_2$Ti$_2$O$_7$ single crystals at room temperature. Structural and compositional analysis along with absorption measurement confirms single crystalline phase of all samples. Room temperature polarized Raman measurements were performed on crystals in non-resonant and two different resonant conditions by using six different laser excitation lines. Lorentzian model fitting analysis is performed on all measured spectra in order to identify the difference in the Raman scattering cross-section in resonant and non-resonant conditions. Variations in the fitting parameters on account of different polarization configurations and crystallographic orientations has helped identifying their symmetry if present. Several possible scattering pathways are discussed in order to qualitatively explain the anomalous scattering results in Ho$_2$Ti$_2$O$_7$.  
\end{abstract}
\maketitle

\section{INTRODUCTION}

Rare-earth metal titanates, RE$_2$Ti$_2$O$_7$, belong to the family of geometrically frustrated magnetic insulators crystallizing in the pyrochlore structure.\cite{Ramirez,GREEDAN} Due to the existance of numerous exotic ground states resulting from competing interactions among spin and orbital degrees of freedom, pyrochlores have drawn enormous interest from the scientific community in recent years. In particular, Ho$_2$Ti$_2$O$_7$ (HTO), a prototype spin-ice system has been estabilished as a weak ferromagnetically frustrated pyrochlore with dominating spin dipolar interaction and competing antiferromagnetic superexchange interaction between neighboring Ho$^{3+}$ ions.\cite{Bramhartog,BramField} The magnetic nature of the ground state and the essence of competing interactions in HTO has been studied extensively at low temperatures through neutron scattering, specific heat capacity, and muon-spin resonance measurements, indicating the existence of a magnetic phase with short-range spin ice correlations due to the incompatibility of local and global symmetries.\cite{Bramhartog,Harrisbramwell,Ramirez}

Raman spectroscopy provides another technique to investigate disorder, phonon anharmonicity, phonon-spin and phonon-crystal field interactions in pyrochlore-type materials. The inelastic scattering of incident photons of fixed polarization due to spins and crystal field coupled phonons allow us to deduce information about the crystal anisotropy, nature of coupling interactions and dynamical characteristics of ground eigenstates. Several temperature dependent Raman spectroscopic studies have been performed, underlining intriguing vibrational band features in rare-earth pyrochlores. However, there have been inconsistencies about their origins in existing literature. Spin-ice pyrochlore Dy$_2$Ti$_2$O$_7$, in addition to typical six Raman modes belonging to the pyrochlore family, has shown several weak modes at the lower and higher frequency ends.\cite{Bi,Lummen,Mkaczka} Their temperature dependent infrared studies suggested strong spin-phonon coupling and the intrinsic charge localization was proposed to result from the nearest neighbour ferromagnetic interaction in a geometrically frustrated configuration of Dy$_2$Ti$_2$O$_7$ spin ice.\cite{Bi} However, the intensity profile of polarization and temperature dependent Raman measurements on Dy$_2$Ti$_2$O$_7$, when combined with the absorption and luminescence spectra, indicate the crystal field transition between stark-split levels.\cite{Lummen,Mkaczka} Another non-magnetic pyrochlore Lu$_2$Ti$_2$O$_7$ also shows similar characteristics in terms of phonon lineshape and locations in the temperature dependent Raman studies. This insinuates that the origin of such features could be non-magnetic and beyond any particular crystal-field effects. Therefore, instead of spin-phonon coupling or crystal-field transitions, it was proposed that either second-order Raman scatterings or infrared and silent modes rendered Raman activity due to lowering of local symmetry could result in additional features.\cite{Saha,Vandenborre} Spectroscopic results for spin-liquid pyrochlore Tb$_2$Ti$_2$O$_7$ implies unusually strong crystal field-phonon coupling along with phonon-phonon anharmonic interaction and small spin-phonon coupling as the possibe origin of atypical features in the spectra.\cite{Sanjuan} Another geometrically frustrated spin-glass like pyrochlore Y$_2$Ru$_2$O$_7$ has been reported to possess strong spin-phonon coupling through their temperature dependent infrared and Raman measurements.\cite{BAE,Lee}. While all such pyrochlore studies have shown many common and few unusual spectral features, none of them have cited any of the resonant Raman modes in their temperature dependent inelastic or quasi-elastic scattering experiments. By tuning the incident photon energy near resonance of localized RE$^{3+}$ atomic levels, one can probe the system into different intermediate eigenstate and track down changes in the scattering phenomena, allowing us to examine the nature of coupling between phonons and other degrees of freedom.  In this article, room temperature polarized Raman spectroscopy is performed on HTO single crystals (SC) with (111) and (001) orientations, probed by multiple incident laser excitations near and far from the resonant energies of localized Ho$^{3+}$ atomic levels. It aims to investigate the remarkably different scattering cross-sections in the resonant and non-resonant conditions and attempts to describe it qualitatively on account of key roles played by other coupling interactions such as spin-exchange, magnetic excitations and crystal field.

\section{EXPERIMENTAL DETAILS}

The single-crystal samples of HTO were grown using the optical floating-zone method. The starting materials Ho$_{2}$O$_{3}$ and TiO$_{2}$ powders were mixed in a stoichiometric ratio and then annealed in air at 1450$^{\circ}$ for 40 h before growth in an image furnace. The growth was achieved with a pulling speed of 6 mm/h under 5 atm oxygen pressure. The crystals were oriented by Laue back diffraction. The structural and compositional analysis on samples were performed by Oxford diffraction Xcaliber2 KMW150CCD and JEOL 7401 FE-SEM with EDAX Genesis XM4 spectroscopy respectively in order to verify the growth integrity and inspect for possible stoichiometric imbalance. X-ray diffraction spectra confirms the cubic symmetry of crystals (\textit{a=b=c=}10.1~A$^{\circ}$; \; $\alpha$ =$\beta$= $\gamma$ =90$^{\circ}$) and no indication of impurity phases has been found. EDS compositional analysis as well confirms 1:1 stoichiometric ratio between Ho and Ti atoms.

The room temperature polarized and unpolarized Raman spectra were measured using Horiba JY LabRam HR800 Raman spectrograph in the back scattered geometry which was coupled to three lasers to supply excitation wavelengths at 785 nm, 633 nm, 514 nm, 488 nm, 458 nm and 364 nm. It uses appropriate bandpass and edge filters to couple the laser beam into the optical axis of Olympus BX30M microscope, equipped with 50x objective and eventually filters out the scattered laser light before the Raman signal enters the spectrograph. LabRam HR800 was equipped with 600 and 1800 lines/mm gratings providing resolution of about 2--3 cm$^{-1}$ in the measurement region. The grating stabilized diode laser providing 785 nm laser excitation was operated at 80 mW (15 mW at the sample) whereas the Melles-Griot 633 nm Helium-Neon laser was operated at 17 mW output power (6 mW at the sample). Coherent I-308 argon ion laser system providing 514 nm, 488 nm, 458 nm and 364 nm laser excitation lines was operated at about 20--30 mW of average power output.

The room temperature absorption measurements on polished HTO SC were performed using an Ocean Optics USB2000 spectrometer in the range of 10,000--29,000 cm$^{-1}$ (345--1000 nm). The spectrograph was collected using 600 lines/mm grating, had 25 micron entrance slit-width giving a spectral resolution of about 1.5 nm FWHM (Full Width at Half Maximum) in the measurement region. 

\section{MEARUREMENTS AND RESULTS}

\begin{figure}
\centering
\includegraphics[width= 3.5 in,height=3.5 in,keepaspectratio]{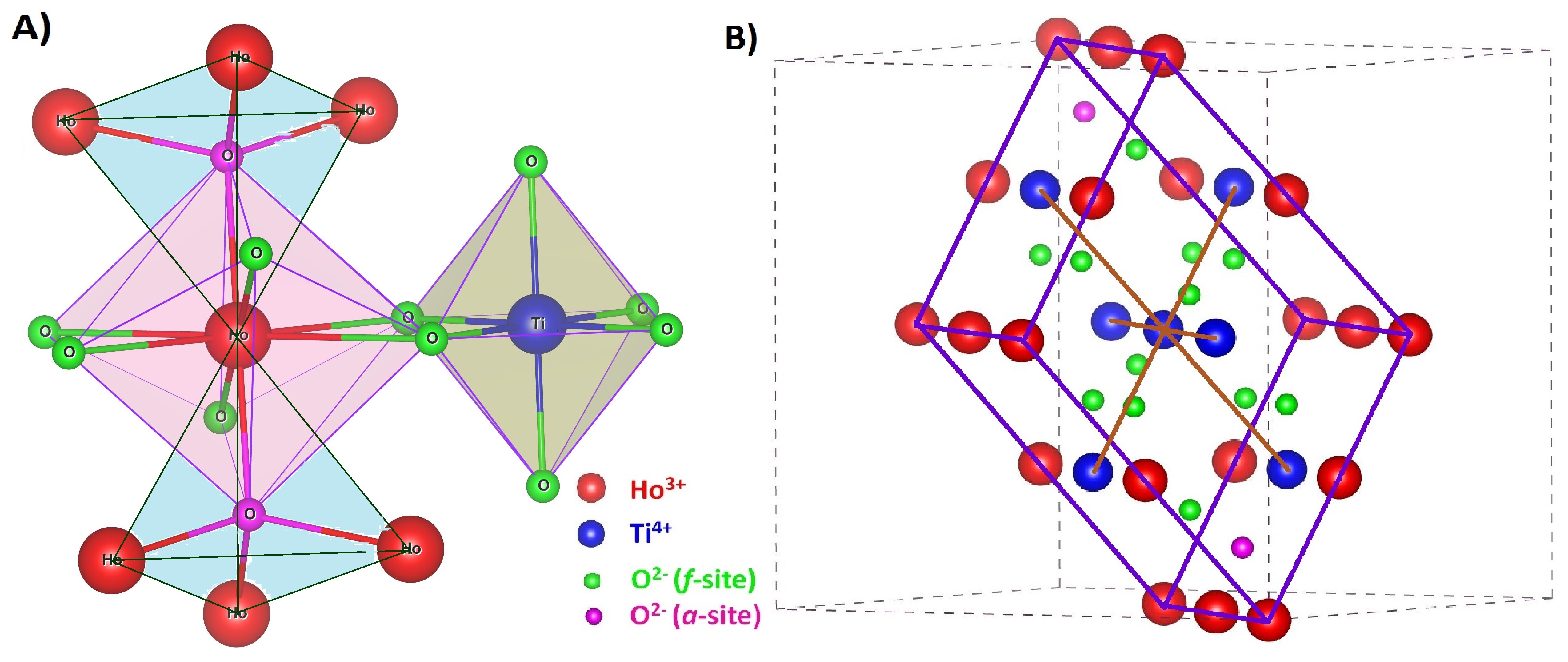}
\caption{\label{HTO}(Color online) 
Crystal structure of HTO showing A) distorted cubic and octahedral crystal field around Ho$^{3+}$ and Ti$^{4+}$ ions respectively. B) Primitive unit cell showing interpenetrating FCC unit cells formed by both Ho$^{3+}$ and Ti$^{4+}$ ions. Primitive cell contains 4 Ho$^{3+}$, 4 Ti$^{4+}$, 12 \textit{f}-type and 2 \textit{a}-type O$^{2-}$ ions.}
\end{figure}

HTO has a cubic structure (lattice parameter 10.1~A$^{\circ}$), crystallizing in Fd$\bar{3}$m space group with eight formula units in a unit cell. The eight-coordinated Ho$^{3+}$ ions are located at \textit{16c} sites whereas six-coordinated Ti$^{4+}$ ions are located at \textit{16d} sites as shown in panel A) of Fig.~\ref{HTO}, both forming separate networks of corner sharing teterahedra. The oxygen anions of one kind occupy \textit{48f} sites coordinating with two Ho$^{3+}$ and two Ti$^{4+}$ ions whereas the oxygen anions of other kind occupy \textit{8a} sites being tetrahedrally coordinated with four Ho$^{3+}$ ions, also shown in panel A) of Fig.~\ref{HTO}.\cite{GardnerJ} Based on lattice parameters, atomic Wyckoff positions and lattice symmetry as shown in panel B) of Fig.~\ref{HTO}, the entire set of degrees of freedom is expressed in terms of following irreducible point group representation at the center of the Brillouin zone.

\vspace{-5pt}
\vspace{-5pt}

\vspace{-5pt}
\begin{equation}
\begin{aligned}
\Gamma_{3N}= {} & \textcolor{red}{1A_{1g}+2E_{g}+12T_{2g}} \textcolor{blue}{+ 24T_{1u}}\\
& \textcolor{green}{+ 6T_{1g}+3A_{2u}+6E_{u}+12T_{2u}}
\end{aligned}
\end{equation}
\vspace{-5pt}

\noindent Here \textit{N} denotes total number of atoms in the primitive cell which is 22 (4 Ho, 4 Ti, 12 \textit{f}-type and 2 \textit{a}-type O), also shown in panel B) of  Fig.~\ref{HTO}. All modes in red color represent raman active modes (total 15 modes) while all in blue are infrared active modes (total 24 modes including 3 acoustic modes). The rest of the 27 modes in green are optically inactive  modes. \textit{E$_{g}$} and \textit{T$_{2g}$} modes are double and triple degenerate, respectively, so one should be able to locate a total of 6 distinct first order Raman active modes in any unpolarized Raman measurement. In addition, all Raman active vibrational modes consist of oxygen atom dynamics only (1 \textit{A$_{1g}$}, 1 \textit{E$_{g}$} and 3 \textit{T$_{2g}$} modes from oxygen anion at \textit{48f} sites, and 1 \textit{T$_{2g}$} mode from oxygen anion at \textit{8a} sites). As both cation sites possess inversion symmetry, any cation vibrational mode should be Raman inactive.


\subsection{Non-resonant unpolarized Raman scattering}
\begin{figure}
\centering
\includegraphics[width= 3.5 in,height=3.5 in,keepaspectratio]{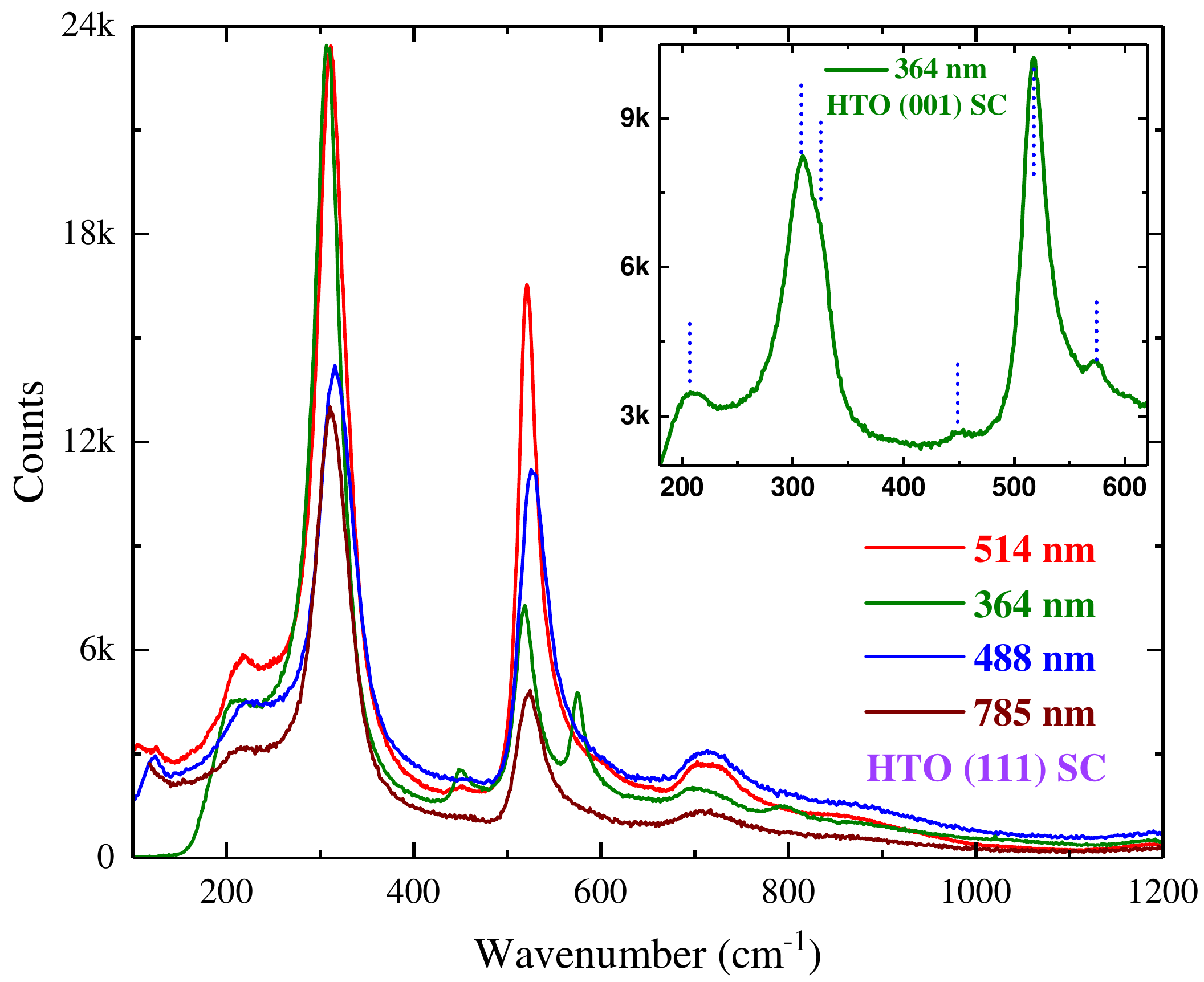}
\caption{\label{unpol}(Color online) 
Unpolarized Raman spectra of (111) HTO SC measured at several laser excitation lines. Inset showing the zoomed-in spectra for (001) HTO SC measured with 364 nm line.}
\end{figure}

Unpolarized Raman measurements were performed on HTO SC with (111) and (001) orientations using 785 nm, 514 nm, 488 nm and 364 nm laser excitation lines and results are shown in Fig.~\ref{unpol}. There have been several experimental and first-principle studies on vibrational properties of HTO SC showing inconsistencies with one another and variations depending on the sample quality.\cite{Lummen,Mkaczka,KUMAR,Ruminy,Kushwaha} Although, Raman spectra for (001) and (111) crystals were very similar, using multiple laser lines on two differently oriented crystals has helped us clearly identify all the six first order Raman active modes.  More details on their symmetries are discussed later in the polarized Raman measurements section.

All Raman spectra on (111) HTO SC in Fig.~\ref{unpol} show consistency in terms of mode locations and relative intensities except for the 364 nm laser line. The spectrum for 364 nm excitation highlights two additional Raman modes clearly resolved above the background at 450 cm$^{-1}$ and 570 cm$^{-1}$, which are not obvious in other spectra, and they have not experimentally observed previously.\cite{Lummen,Mkaczka} Inset plot shows 364 nm Raman spectrum for (001) HTO SC where modes at 310 cm$^{-1}$ and 330 cm$^{-1}$ are partially resolved, not obvious in other specta due to thermal broadening of phonons at room temperature. Based on prior studies, modes at 330 cm$^{-1}$ and 520 cm$^{-1}$ have been assigned \textit{E$_{g}$} and \textit{A$_{1g}$} symmetries respectively whereas modes with \textit{T$_{2g}$} symmetry are located at 220 cm$^{-1}$, 310 cm$^{-1}$, 450 cm$^{-1}$ and 570 cm$^{-1}$.\cite{Lummen,Mkaczka} The mode at 720 cm$^{-1}$ has \textit{A$_{1g}$} symmetry, which results due to higher order scattering process, however common in rare-earth pyrochlores, details on such higher order processes are pending. 

\vspace{-5pt}
\vspace{-5pt}

\subsection{Absorption spectra and resonant polarized Raman scattering}

\begin{figure*}
\centering
\includegraphics[width=\textwidth]{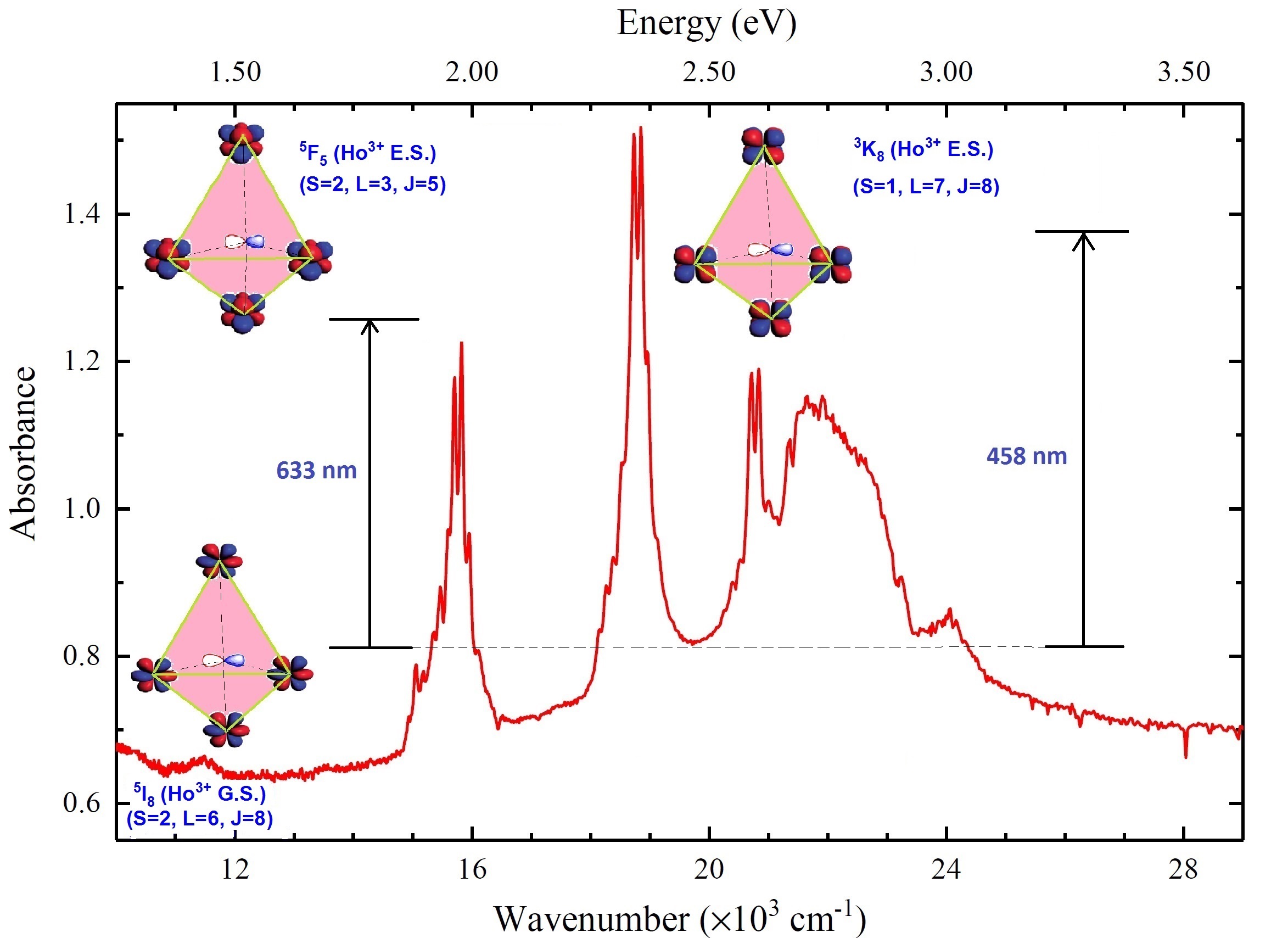}
\caption{\label{Absorption}(Color online) Room temperature absorption spectrum of (111) HTO SC showing transitions from ground state to higher spin-orbit coupled excited states with sharp peaks showing transitions to multiple crystal field levels. Two specific Raman laser lines in the transition schematics indicate the difference in orbital shapes (representative) of Ho$^{3+}$ ions in tetrahedra network, possibly changing orbital overlap and exchange interaction upon excitation.}
\end{figure*}

Room temperature absorption measurements were performed on (111) and (001) HTO SC to locate Ho$^{3+}$ atomic levels. This was essential to find appropriate Raman laser lines to probe the system into resonance. Such intra-atomic resonance alters the nature of spin-orbit coupling and exchange interaction between localized spins, which helps us to understand the mechanism of phonon scattering mediated by magnons and spin-disorders.\cite{Merlin,Guntherodt,Zeyher,Sugai,Chubukov,Rubhausen,Blumberg} Fig.~\ref{Absorption} shows the absorption spectrum for (111) HTO SC where transitions between several excited states and their crystal-field (CF) split levels have been displayed. The spectrum looks in good agreement with previously reported absorption results.\cite{MACALIK} Based on proposed atomic energy level scheme,\cite{Martin,Carnall,dieke} identification of several relevant transitions from the $^{5}I_{8}$ ground state (S=2, L=6, J=8) to other excited states has been made. The 633 nm laser line excites the system into $^5F_5$ state (S=2, L=3, J=5) whereas the 458 nm laser line excites into $^3K_8$ spin-orbit manifold (S=1, L=7, J=8). These spin-orbit coupled states consist of a finite number of closely spaced crystal-field energy levels as seen in the absorption spectrum as sharp spikes. Since, the excited states with different orbital and total angular momentum affect the overlap integral with neighboring oxygen anions, and the exchange interaction between Ho$^{3+}$ ions in the tetrahedral network, significant changes in the resonant Raman scattering cross section are expected.

\begin{figure}
\centering
\includegraphics[width= 3.5 in,height=3.5 in,keepaspectratio]{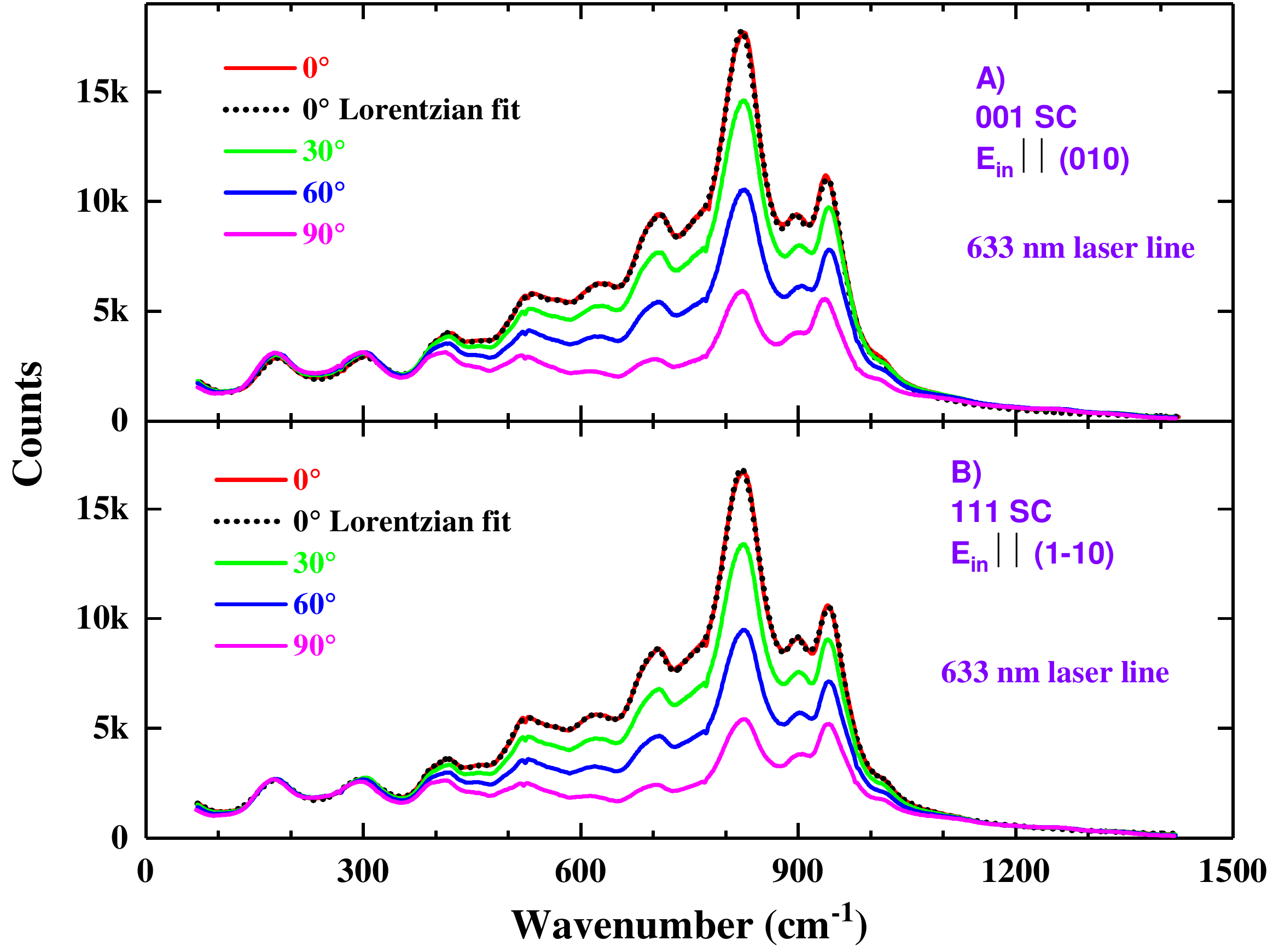}
\caption{\label{633}(Color online) 
Polarized Resonant Raman spectra of A) (001) and B) (111) HTO SC using 633 nm laser line in back-scattered geometry. (Analyzer axis $\parallel \vec{E}_{in}$ = 0$^{\circ}$, Analyzer axis $\perp \vec{E}_{in}$ = 90$^{\circ}$)}
\end{figure}

Polarized resonant Raman measurements have been performed on (001) and (111) HTO SC in back-scattered geometry for several polarizer-analyzer configurations using 633 nm laser line as shown in Fig.~\ref{633}. Panel A) shows spectra for (001) SC with $\vec{E}_{in}$ parallel to [010] axis while panel B) shows spectra for (111) SC with $\vec{E}_{in}$ parallel to [1$\bar{1}$0] axis. The analyzer transmission axis is rotated in 30$^{\circ}$ steps for both measurements where 0$^{\circ}$ spectrum represents $\vec{E}_{in}$ being parallel to the analyzer transmission axis. All spectra have been fitted with Lorentzian model using HORIBA Scientific’s LabSpec 6 software platform and 0$^{\circ}$ fitted curve is included for both crystals. Similar spectra and their fittings with $\vec{E}_{in}$ parallel to the other perpendicular axis [11$\bar{2}$] and at 45$^{\circ}$ away from [11$\bar{2}$] for (111) SC and then $\vec{E}_{in}$ parallel to [100] and [110] for (001) SC were also collected and shown in the supplementary section in Fig.~\ref{111633f} and Fig.~\ref{001633f}, respectively. In addition to few weak ones, all collected spectra for every polarizer-analyzer configuration show twelve distinguished anomalous modes at 180 cm$^{-1}$, 300 cm$^{-1}$, 390 cm$^{-1}$, 420 cm$^{-1}$, 520 cm$^{-1}$, 620 cm$^{-1}$, 680 cm$^{-1}$, 710 cm$^{-1}$, 750 cm$^{-1}$, 820 cm$^{-1}$, 900 cm$^{-1}$ and 945 cm$^{-1}$ with extremely relaxed symmetry. All modes above 300 cm$^{-1}$ show monotonic decrease in the oscillator strength with varying depolarization ratio. While comparing 0$^{\circ}$ (parallel polarized) with 90$^{\circ}$ (perpendicular polarized) spectrum, none of the modes including weak ones completely disappear or appear for either of the (001) and (111) HTO SC for any of the six crystallographic measurement directions of the incident polarization $\vec{E}_{in}$. Detailed analysis on their behaviors are discussed in the next section.  

\begin{figure}
\centering
\includegraphics[width= 3.5 in,height=3.5 in,keepaspectratio]{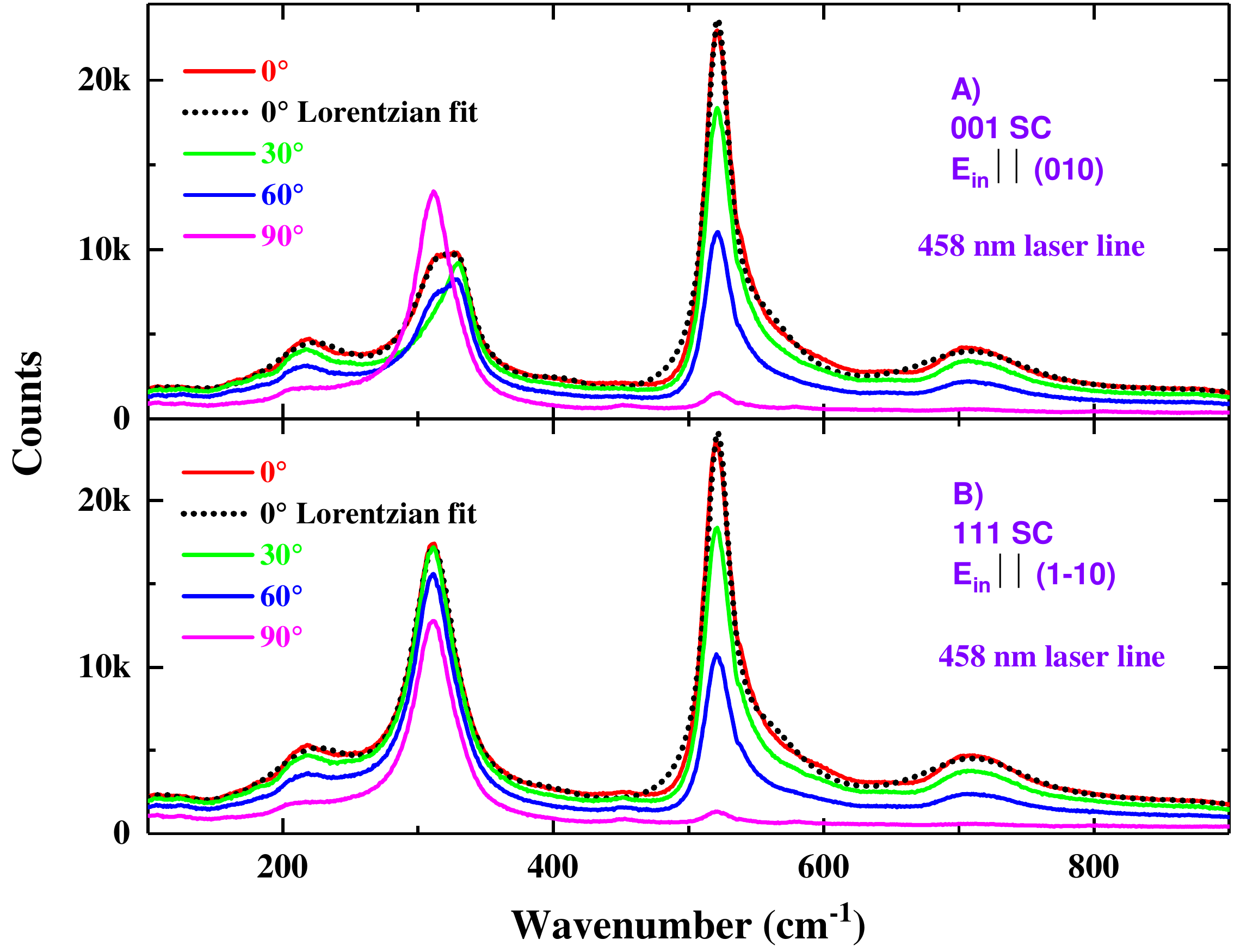}
\caption{\label{458_1}(Color online) Polarized resonant Raman spectra of A) (001) and B) (111) HTO SC using 458 nm laser line in back-scattered geometry at lower frequency region. (Analyzer axis $\parallel \vec{E}_{in}$ = 0$^{\circ}$, Analyzer axis $\perp \vec{E}_{in}$ = 90$^{\circ}$)}
\end{figure}

\begin{figure}
\centering
\includegraphics[width= 3.5 in,height=3.5 in,keepaspectratio]{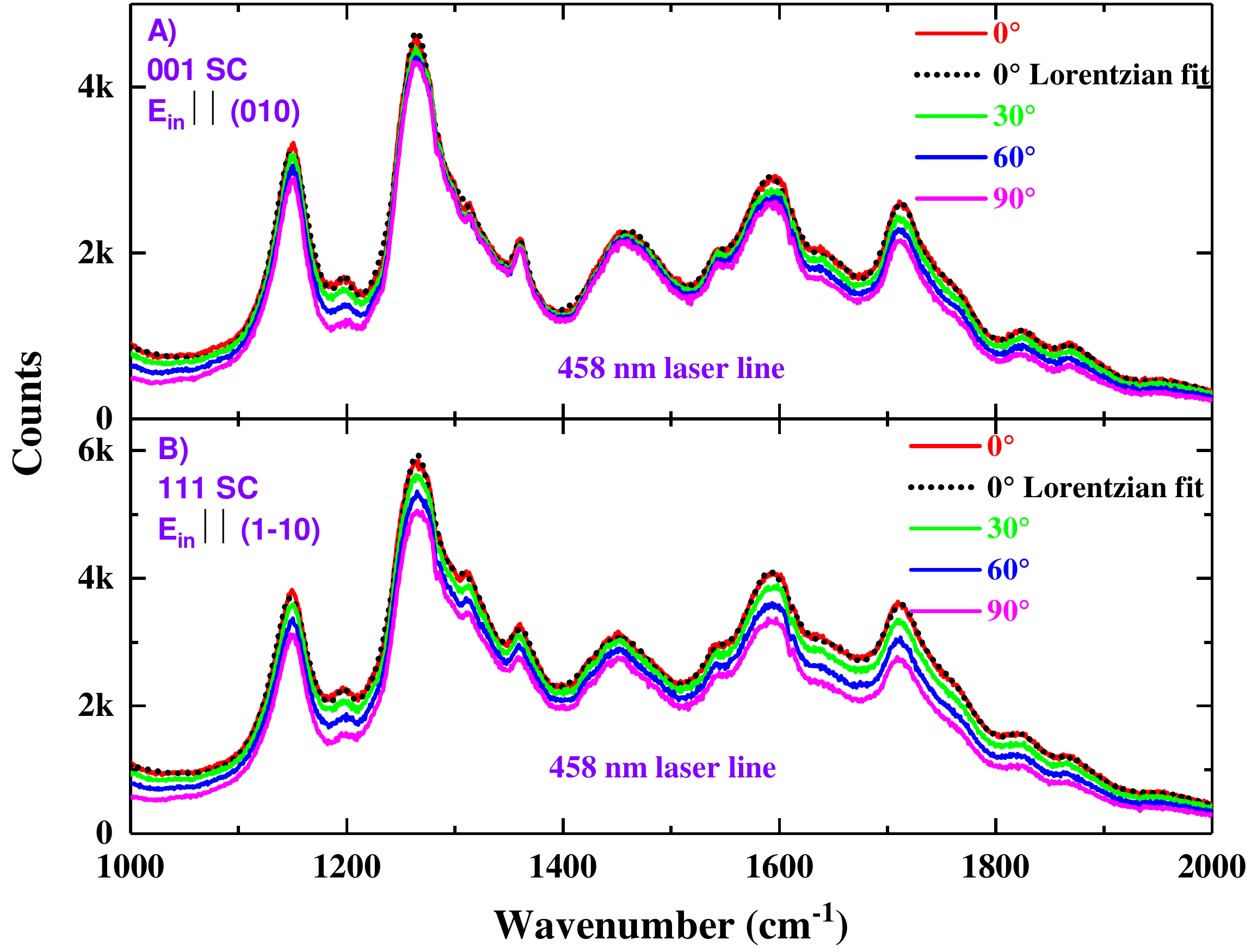}
\caption{\label{458_2}(Color online) Polarized resonant Raman spectra of A) (001) and B) (111) HTO SC using 458 nm laser line in back-scattered geometry at higher frequency region. (Analyzer axis $\parallel \vec{E}_{in}$ = 0$^{\circ}$, Analyzer axis $\perp \vec{E}_{in}$ = 90$^{\circ}$)}
\end{figure}

In order to further investigate such anomalous scattering phenomena, another resonant laser line of 458 nm was used on the same samples under identical experimental conditions. Measurement results are shown in Fig.~\ref{458_1} for low frequency region (100--900 cm$^{-1}$) and in Fig.~\ref{458_2} for high frequency region (1000--2000 cm$^{-1}$) along with 0$^{\circ}$ spectrum fitted using Lorentzian model. The low frequency spectra in Fig.~\ref{458_1} looks very similar to non-resonant Raman spectra as seen with other laser lines in Fig.~\ref{unpol}. Polarization selection rules for Raman scattering for the Fd$\bar{3}$m space group suggest that in back-scattered configuration for (001) SC, the perpendicular polarized output intensity should only contain \textit{T$_{2g}$} modes while parallel polarized output intensity profile should contain both \textit{A$_{1g}$} and \textit{E$_{g}$} modes. Comparing with panel A) in Fig.~\ref{458_1}, the peak at 310 cm$^{-1}$ in 90 deg spectra must have \textit{T$_{2g}$} symmetry. Similarly, the 520 cm$^{-1}$ and 330 cm$^{-1}$ modes have \textit{A$_{1g}$} and \textit{E$_{g}$} symmetry, respectively. Panel B) of Fig.~\ref{458_1} does not resolve the symmetry of modes as clearly since the Raman tensor is transformed with respect to another set of basis vectors, which has multiple non zero off-diagonal elements and hence, intensity profiles are mixed. However one can see first order Raman modes with some selectivity in this frequency range. 

While the low frequency Raman spectra show expected selectivity among fundamental Raman modes, spectra in the high frequency region (Fig.~\ref{458_2}) show multiple new modes with extremely relaxed selectivity. Fitting results identify at least twelve anomalous modes with negligible selectivity with respect to the incident polarization $\vec{E}_{in}$ or relative orientation of the analyzer transmission axis. These modes are located around 1150 cm$^{-1}$, 1200 cm$^{-1}$, 1260 cm$^{-1}$, 1310 cm$^{-1}$, 1360 cm$^{-1}$, 1460 cm$^{-1}$, 1540 cm$^{-1}$, 1590 cm$^{-1}$, 1650 cm$^{-1}$, 1710 cm$^{-1}$, 1825 cm$^{-1}$ and 1870 cm$^{-1}$. Many of these modes are better resolved and show more sensitivity to analyzer orientation in panel B) for (111) HTO SC measurements. Measurements are repeated for six different crystallographic directions of the incident polarization $\vec{E}_{in}$ in (001) and (111) HTO SC. Those spectra are included in the supplementary section in Fig.~\ref{111458}, Fig.~\ref{111458f}, Fig.~\ref{001458} and Fig.~\ref{001458f}. They are quite similar in terms of mode locations and relative intensity as analyzer is rotated.

Other similarly grown single crystals from the pyrochlore family were also measured under identical experimental conditions in order to exclude the possibility of experimental set-up induced artifacts such as unstable or nonmonochomatic laser line source, filter leakage, damaged polarizer-analyzer optic axis and optical grating induced errors. Measurement results are shown in Fig.~\ref{all} for two different laser lines. The non-resonant Raman results for 514 nm is included in the supplementary section in Fig.~\ref{514nonpol}, which looks very similar in terms of mode location line shape and relative intensity profile for all rare-earth pyrochlores. In fact 633 nm spectra in panel A) and 458 nm spectra in panel B) look very similar to 514 nm spectra for all pyrochlores except Ho$_2$Ti$_2$O$_7$. This observation without any ambiguity indicates the significant change in Raman scattering cross-section under resonance condition for HTO pyrochlore. Measurements have been repeated on several batches of pyrochlore samples with different sample sizes and roughnesses. All spectra seem to be independent of sample variation. The Raman spectra using 633 nm line were collected at several filter-attenuation settings for (111) HTO SC as shown in supplemental section in Fig.~\ref{Intensity}. Output signal at two major peaks (820 cm$^{-1}$ and 945 cm$^{-1}$) has been analyzed and shown in the inset graph. A linear trend between varying input intensity and corresponding output intensity suggests the absence of any non linear processes such as multiphoton absorption, stimulated Raman, thermal effects on sample itself or non-linear optical effects from the optical components in the set-up. Prolonged exposure of samples to the laser excitation does not result in any obvious spectral or background changes ruling out the possibilities of any sample surface damage or well cited fluorescence background issues with resonant Raman technique.\cite{Yaney,Kostamovaara,McCain,Mazilu,smithz,Matousek} These intriguing modes resulting from the inelastic scattering of incident photon could be an outcome of crystal vibrations interacting with spin disorder or non-trivial spin-correlations in the excited state or crystal field.\cite{Merlin,Guntherodt,Zeyher,GRUNBERG,SCHLEGEL,TEKIPPE} As discussed in the next section, several rare-earth magnetic compounds exhibit such scattering phenomena and, when the resonance conditions are met, do receive resonance enhancement of several orders of magnitude. On account of similar previous research studies, the next section provides a qualitative discussion on such possibilities and the future research problems that need to be further explored. Moreover, rare-earth atomic level database suggested that resonance conditions were not possible to achieve with any accessible laser lines for all other available pyrochlores.\cite{Weber,CAVALLI,Martin,Carnall,dieke} 

\begin{figure}
\centering
\includegraphics[width= 3.5 in,height=3.5 in,keepaspectratio]{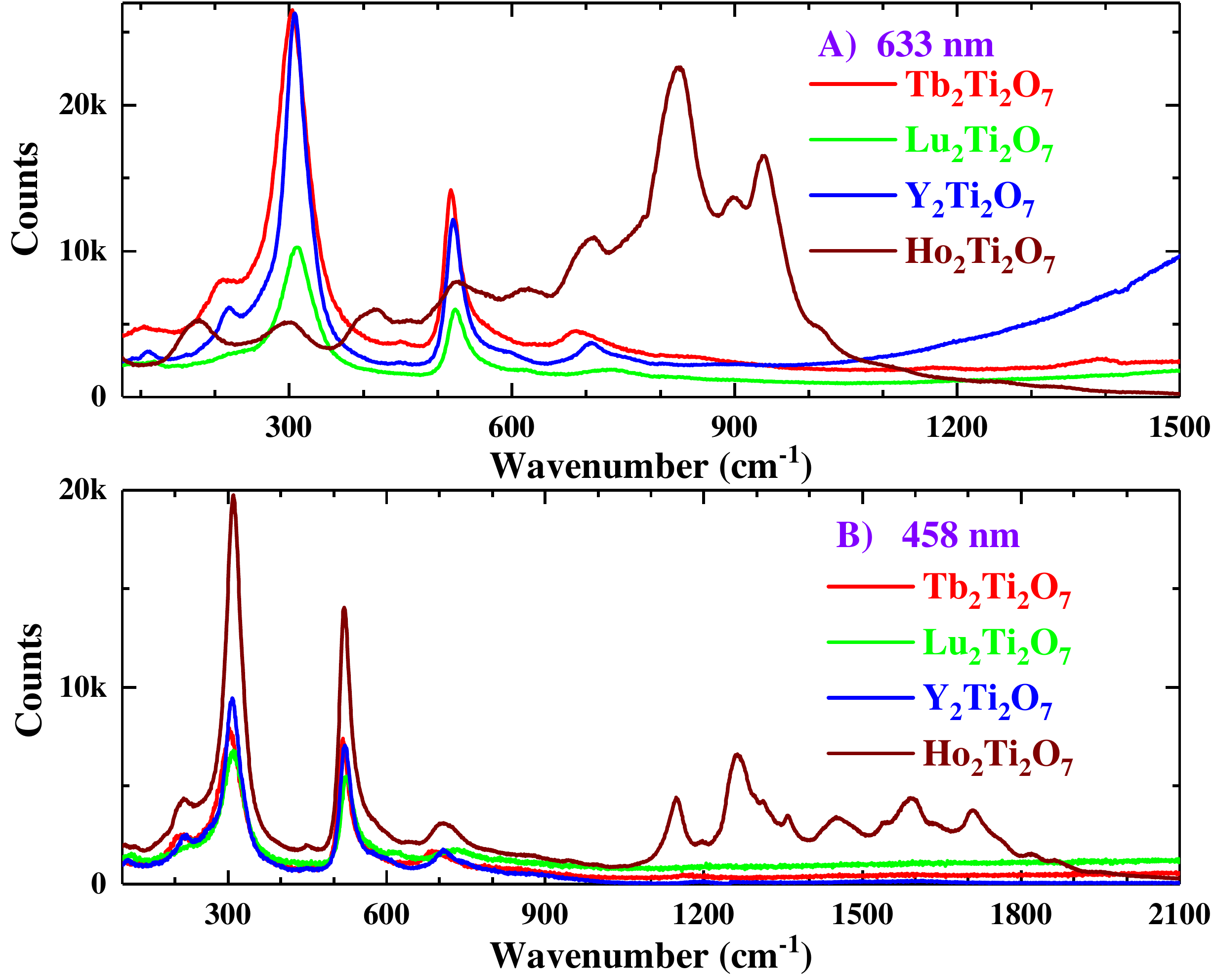}
\caption{\label{all}(Color online) Unpolarized Raman measurements on several (111) rare-earth pyrochlore single crystals using A) 633 nm B) 458 nm. Both spectra for Ho$_2$Ti$_2$O$_7$, with remarkably different Raman cross-section as compared to other pyrochlores, show resonance effect on scattering phenomena.}
\end{figure}

\section{ANALYSIS AND DISCUSSION} 

\begin{figure}
\centering
\includegraphics[width= 3.5 in,height=3.5 in,keepaspectratio]{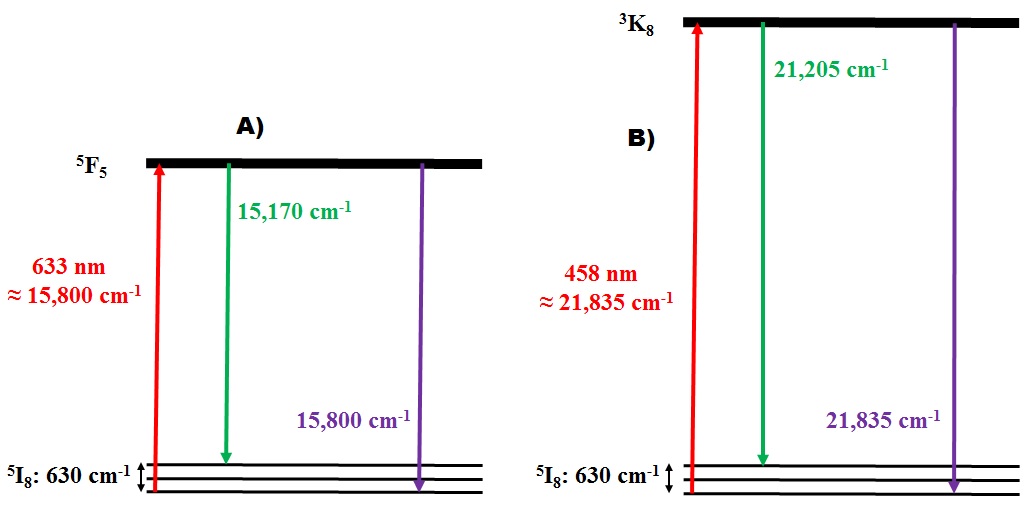}
\caption{\label{electronicbandstructure}(Color online) A representative excitation-relaxation schematic for A) 633 nm (15,800 cm$^{-1}$) B) 458 nm (21,835 cm$^{-1}$) laser lines. Green and purple colored relaxations show the lower and upper limits on the fluorescent photons, respectively.}
\end{figure}

Several Raman studies on rare-earth pyrochlores have reported spectral features originating due to the transition between the crystal field split states of $^{5}I_{8}$ ground state manifold in the non-resonant condition.\cite{Mkaczka,Lummen,Sanjuan} Any electronic transition within the same $\left|L,J\right\rangle$ states are generally dipole forbidden and, unless CF states are coupled with other degrees of freedom such as phonons or spin disorder, such transitions do not satisfy the momentum conservation rule. This explains their extremely weak intensity pattern in comparison to Raman modes.\cite{Mkaczka,Lummen,Sanjuan} In HTO, the overall CF splitting of $^{5}I_{8}$ ground state is about 630 cm$^{-1}$\cite{Rosenkranz} and, 633 nm (15,800 cm$^{-1}$) laser line excites electrons from the ground state to the $^{5}F_{5}$ states. If excited electrons relax back to any of the $^{5}I_{8}$ CF states, the energy of such fluorescent output photons should lie in the range of 15,170--15,800 cm$^{-1}$ as shown in the panel A) of Fig.~\ref{electronicbandstructure}. One must notice that the strong Raman Stokes lines around 710 cm$^{-1}$, 820 cm$^{-1}$ and 945 cm$^{-1}$ as shown in Fig.~\ref{633} lie outside this range. Similarly, in the case of 458 nm (21,835 cm$^{-1}$) laser line, electrons get excited into the $^{3}K_{8}$ states and relaxation back to any of the $^{5}I_{8}$ ground states should generate photons in the energy range of 21,205--21,835 cm$^{-1}$ as shown in the panel B) of Fig.~\ref{electronicbandstructure}. All strong Stokes modes between 1150 cm$^{-1}$ and 1870 cm$^{-1}$, shown in Fig.~\ref{458_2}, lie outside the aforementioned energy range. This exercise convinces us that observed spectra in resonant Raman scattering cases as shown in Fig.~\ref{633} and Fig.~\ref{458_2} can not be explained though hot luminescence relaxation mechanism.

\begin{figure*}
\centering
\includegraphics[width=\textwidth]{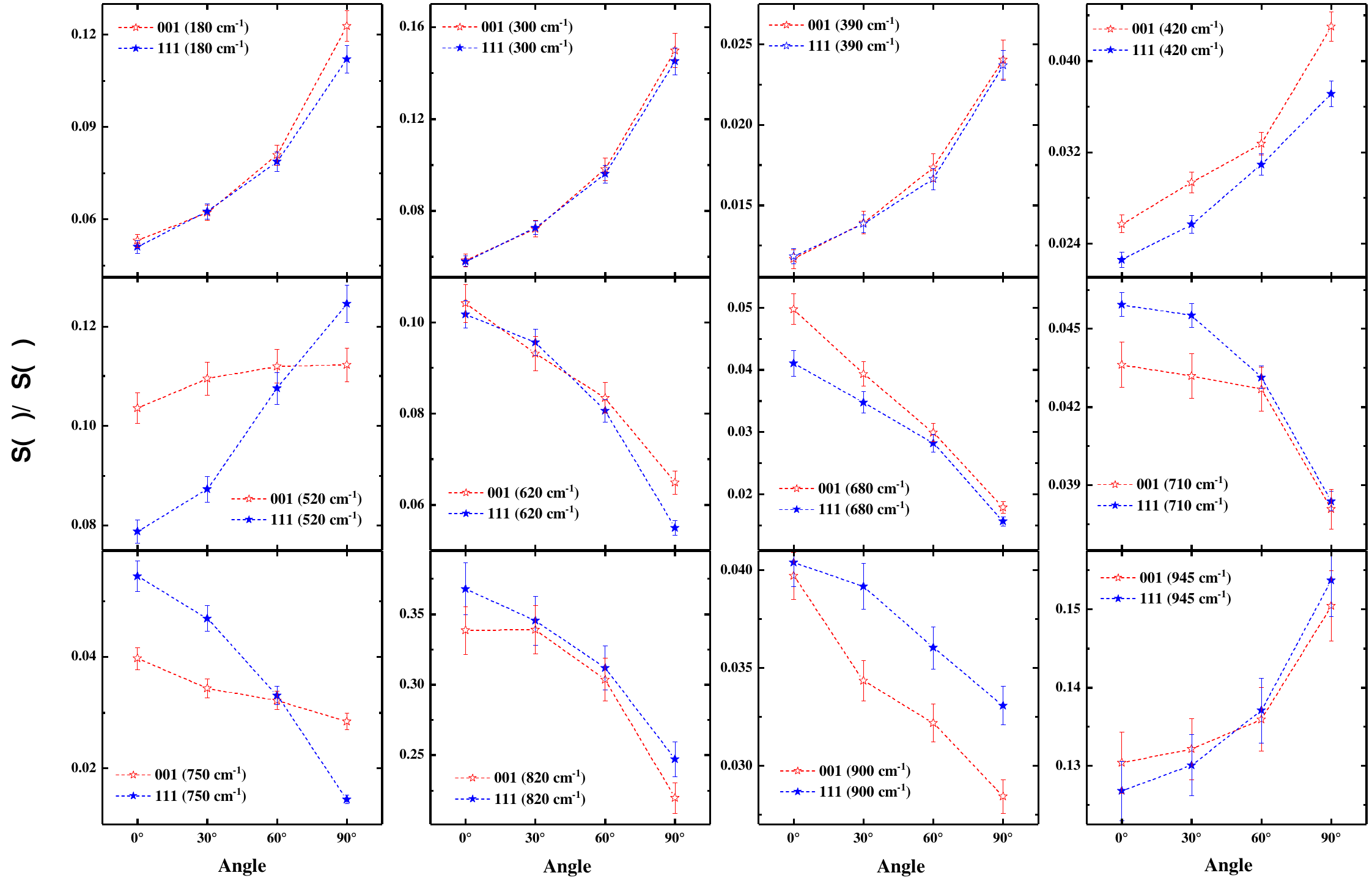}
\caption{\label{ana_633}(Color online) Normalized oscillator strength and the polarization dependence of all strong modes as identified through fitting routine for 633 nm. $\vec{E}_{in}$ $\parallel$ [010] and $\vec{E}_{in}$ $\parallel$ [1$\bar{1}$0] for (001) and (111) HTO SC respectively. \textit{x}-axis denotes the angle between $\vec{E}_{in}$ and analyzer transmission axis.}
\end{figure*}

\begin{figure*}
\centering
\includegraphics[width=\textwidth]{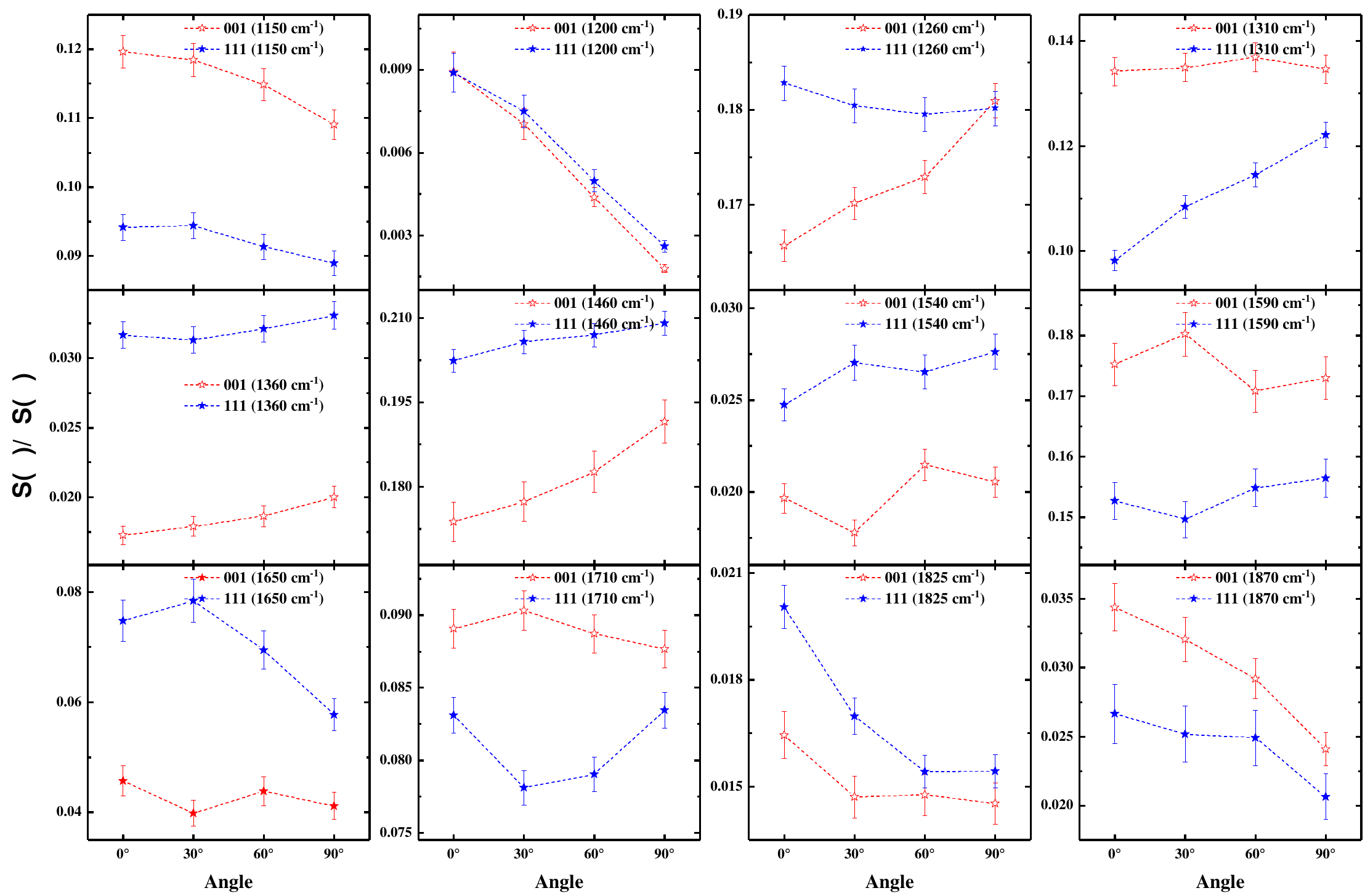}
\caption{\label{ana_458}(Color online) Normalized oscillator strength and the polarization dependence of all strong modes (excluding low energy fundamental modes) as identified through fitting routine for 458 nm. $\vec{E}_{in}$ $\parallel$ [010] and $\vec{E}_{in}$ $\parallel$ [1$\bar{1}$0] for (001) and (111) HTO SC respectively. \textit{x}-axis denotes the angle between $\vec{E}_{in}$ and analyzer transmission axis.}
\end{figure*}

All spectra of HTO under resonance conditions were fitted using Lorentzian model to fully characterize the dynamical parameters of anomalous modes. Results are summarized in Fig.~\ref{ana_633} for 633 nm spectra and in Fig.~\ref{ana_458} for 458 nm spectra, showing the polarization dependence of all strong and well resolved modes for crystals of (001) and (111) orientations. The incident polarization $\vec{E}_{in}$ is parallel to [010] for (001) SC whereas for (111) SC, $\vec{E}_{in}$ is parallel to [1$\bar{1}$0]. The \textit{x}-axis denotes the angle between $\vec{E}_{in}$ and analyzer transmission axis. For both analysis, strength of all modes are normalized with respect to the sum of all oscillator strengths. 

In the case of 633 nm Raman spectra, except few high frequency modes, majority of modes lie in the range where the first order fundamental modes of HTO are theoretically predicted\cite{KUMAR,Ruminy,Kushwaha} and experimentally found if they are optically active\cite{Lummen,Mkaczka}. However, it must be emphasized here that although some of these modes may have similar vibrational frequencies, none of them behave as first order Raman modes when compared to the spectra shown in Fig.~\ref{unpol}. All modes below 300 cm$^{-1}$ show no change in oscillator strength but all higher modes decrease in strength monotonically as the analyzer rotates from parallel polarization to perpendicular polarization configuration. When oscillator strengths are normalized as shown in Fig.~\ref{ana_633}, modes at 180 cm$^{-1}$, 300 cm$^{-1}$, 390 cm$^{-1}$, 420 cm$^{-1}$, 520 cm$^{-1}$ and 945 cm$^{-1}$ seem to prefer perpendicular polarization configuration whereas others at 620 cm$^{-1}$, 680 cm$^{-1}$, 710 cm$^{-1}$, 750 cm$^{-1}$, 820 cm$^{-1}$ and 900 cm$^{-1}$ prefer parallel configuration for both crystals. 

The 458 nm spectral analysis has been conducted separately in two obvious frequency ranges. The low frequency spectra in Fig.~\ref{458_1} show expected trends in terms of symmetry, lineshape and relative intensity profile, just like first order Raman modes shown in Fig.~\ref{unpol}. However, general trends of all modes in higher frquency range as shown in Fig.~\ref{458_2} are remarkably different. Since, these resonance induced anomalous modes are not reported in any previous studies, we have performed detailed spectral analysis for high frequency region and results are shown in Fig.~\ref{ana_458}. While strength of all modes decreases monotonically as analyzer rotates towards perpendicular polarization configuration, the normalized strengths for many modes show very weak polarization dependence. Modes at 1150 cm$^{-1}$, 1200 cm$^{-1}$, 1825 cm$^{-1}$ and 1870 cm$^{-1}$ prefer parallel configuration whereas modes at 1360 cm$^{-1}$ and 1460 cm$^{-1}$ prefer perpendicular polarization configuration for both crystals. While modes at 1540 cm$^{-1}$, 1590 cm$^{-1}$ and 1710 cm$^{-1}$ show no apparent polarization dependence within the error bars for any crystal, modes at 1260 cm$^{-1}$, 1310 cm$^{-1}$ and 1650 cm$^{-1}$ show different behavior for (001) and (111) crystals.  Overall behavior of these phonon modes suggests that while they all show higher strength in parallel configuration, there is certain selectivity or polarization preference among them. Their specifics possibly lie within the details of the polarizability tensor and the involved chemical bonds in the excited states when the resonance condition is met.

In systems like cuprates with strong electron-phonon coupling and rare-earth chalcogenides, phosphates or vanadates in which phonons are known to be coupled with spins and crystal fields, phonon symmetry gets modified and revised selection rules for phonon excitations come into effect.\cite{SugaiShunji,Merlin,Guntherodt,Zeyher} In addition, if one also considers the possibility of higher order Raman processes under the condition of resonance enhancement, a significant change in the single-phonon Raman scattering cross-section could be expected. This would essentially allow the spectrum to exhibit additional features, potentially coming from the entire Brillouin zone (BZ), as long as the momentum conservation is obeyed in the process. Two-phonon peaks could exhibit larger intensity than the single-phonon peaks due to the forbidden selection rule of Raman-inactive modes for scattering.\cite{SugaiShunji} The fact that Ho$_2$Ti$_2$O$_7$ contains 26, 44, 33 and 33 distinct modes of vibrations around $\Gamma$, \textit{L}, \textit{X} and \textit{W} symmetry point respectively, as shown in the supplemental section in Fig.~\ref{BZ}, one can not rule out the possibility of higher order scattering processes being in effect. However, in the absence of details on phonon density of states and vibrational band dispersion inside the Brillouin zone, any prediction on higher order Raman scattering is highly speculative. Using O-18 or Ho-163 isotopes for growth could help identifying the \textit{k}-point origin of these modes and their associated symmetry inside the Brillouin-zone. Nevertheless, higher order phonon modes in many materials have been reported to possess extremely non-Lorentzian lineshape\cite{Gillet,Slawomir,Tenne} while all our spectra for any six different crystallographic directions have been fitted using well defined Lorenzian modes. In addition, anomalous modes in Fig.~\ref{633} and Fig.~\ref{458_2} do not seem to obey the crystal symmetry imposed selection rule and none of the overtones of first order Raman modes are located in the spectra. Although certain rare-earth compounds such as Ytterbium chalcogenides prefer parallel polarized configuration for higher-order scattering, irrespective of the crystal orientation as also observed in our measurements,\cite{VITINS,Vitinsj,MerlinHumphreys} all other spectral features suggest that higher order scattering phenomena may not be the leading factor causing strikingly different Raman cross-section. 

In rare-earth chalcogenides such as EuX (X = O, S, Se, Te), optical phonons from the zone boundary has been reported as the dominant signal in the first-order Raman scattering cross-section. The momentum conservation in such scattering events is accomplished through flipping of electron spins in the spin-disordered paramagnetic phase, resulting into the magnon excitation.\cite{GRUNBERG,SCHLEGEL,TEKIPPE} In principle, this mechanism still allows all the modes to maintain the Lorentzian lineshape, as happens in our case, with just another spin-dependent scaling prefactor affecting the overall intensity profile. The fact that measured scattering intensity in such chalcogenides overlaps with the weighted one-phonon and one-magnon density of states reflects the simultaneous excitation of phonon-magnon quasi-particles.\cite{Merlin,Guntherodt,Zeyher} Similar spin flipping mechanism is recently reported in both spin-ice Titanium pyrochlore where monopole dynamics at high temperature is quantitaively described through possible coupling between crystal-field and optical phonon excitations followed by the phonon-mediated spin-flipping, as observed through quasielastic neutron scattering.\cite{RuminyFennell}. More details about spin density of states in pyrochlore structures could pave the pathway for concurrent excitation of phonon-magnon quasi-particle pair. A systematic temperature dependent resonant Raman scattering experiment could help to further investigate these anomalous modes since the lineshape of modes evolve very differently with temperature if the phonon is coupled with crystal field, or some spin-like degree of freedom, as compared to just simply undergoing anharmonic phonon-phonon interaction.\cite{Mkaczka} On the other hand, a magnetic field dependent resonant Raman studies would allow the tuning of phonon-magnon coupling, resulting into a modified field-dependent spin-flipping mecahnism in the paramagnetic disordered phase. Cooling below the critical temperature of some of the ferromagnetic chalcogenides results into a strong quenching of the scatttering intensity along with a nonlinear peak-shift near critical temperature. Moreover, the symmetry of the spin system dictates the symmetry of the scattered phonon as observed through zone folding of phonon branches over magnetic Brillouin zone.\cite{Snow,Nawrocki,RAY} Finding such key signatures in Ho$_2$Ti$_2$O$_7$ through temperature and field dependent Raman studies will further strengthen the concept of inelastic scattering through concurrent phonon-magnon excitation. In addition, more rare-earth pyrochlores, spin-ice Dy$_2$Ti$_2$O$_7$ in particular, should be tested under resonant conditions and verified if anomalous phonon scattering is correlated with the high temperature monopole dynamics of the frustrated 2-in/2-out spin-ice paramagnetic disordered phase.  

\section{CONCLUSIONS}
Room-temperature polarized Raman measurement were performed on (001) and (111) HTO  SC along with few other rare-earth pyrochlores near resonance and away from resonance conditions. In addition to previously well cited first-order Raman modes, several new phonon modes have been identified. These anomalous phonon modes have strikingly different Raman cross-section from the non-resonant condition. Measurement under identical experimental conditions, but being at different resonant excited state, through separate laser excitations has also resulted into very different Raman cross-section, strongly indicating the role played by coupling interaction with other degrees of freedom such as, spin and crystal-field. Systematic Lorentzian model fitting routine for different polarization configuration and crystallographic orientations has helped in identifying some selectivity among the anomalous phonon modes.

\section{ACKNOWLEDGEMENT}
The authors wish to thank the support from NSF with grant No. DMR-1350002 (Q.H. and H.D.Z.). We also acknowledge Material Characterization Laboratory in the Department of Chemistry \& Biochemistry, Florida State University for providing instrumentation.

\bibliography{HTORaman}

\widetext
\begin{center}
\textbf{\large \textit{Supplemental Material:}Probing Anomalous Inelastic Scattering in Spin-ice Ho$_2$Ti$_2$O$_7$ through Resonant Raman Spectroscopy}
\end{center}

\setcounter{equation}{0}
\setcounter{figure}{0}
\setcounter{table}{0}
\setcounter{page}{1}
\makeatletter
\renewcommand{\theequation}{S\arabic{equation}}
\renewcommand{\thefigure}{S\arabic{figure}}
\renewcommand{\bibnumfmt}[1]{[S#1]}
\renewcommand{\citenumfont}[1]{S#1}
\maketitle

\clearpage

\begin{figure*}
\centering
\includegraphics[width=0.8\textwidth]{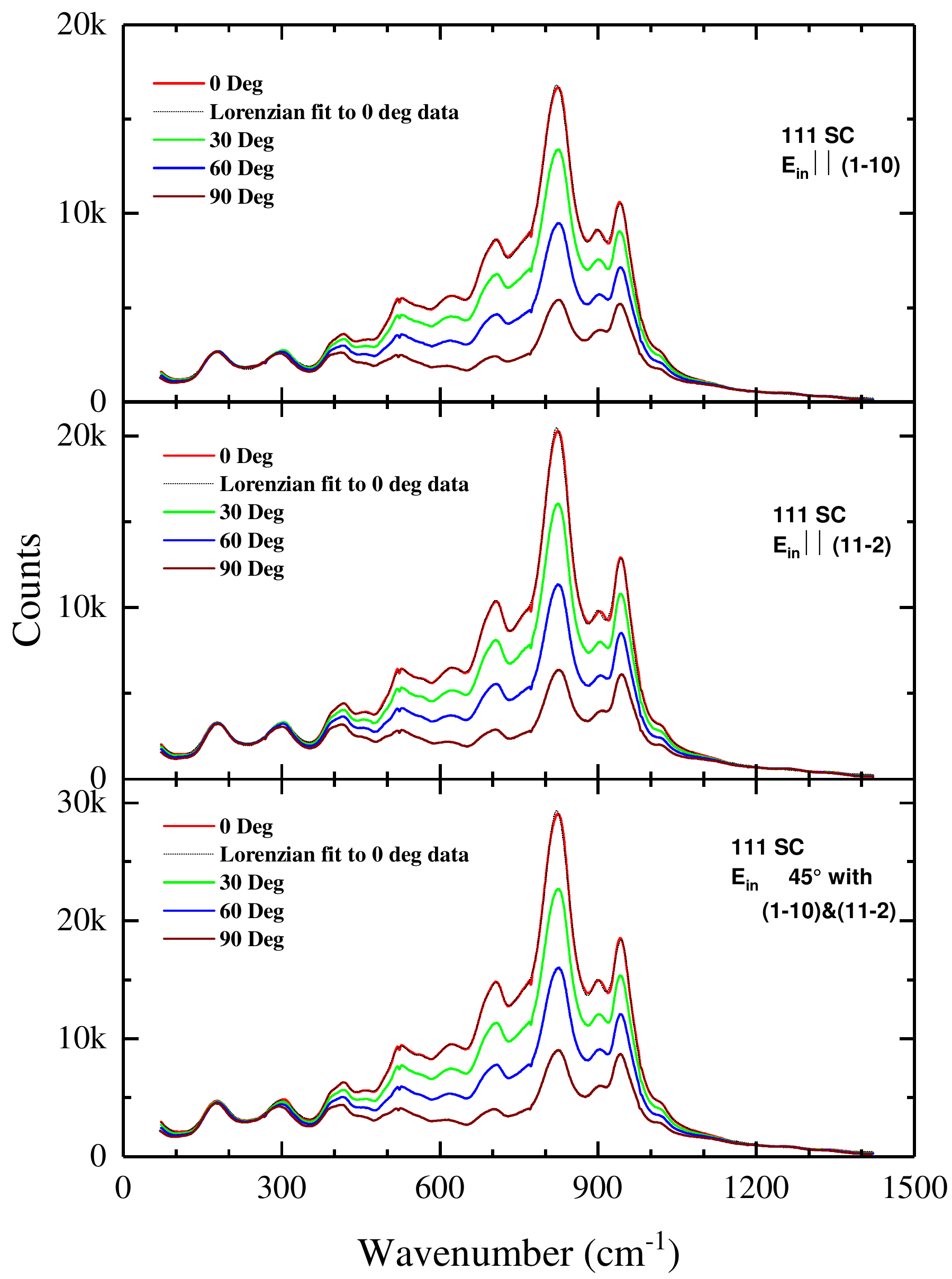}
\caption{\label{111633f}(Color online) 
Polarized resonant Raman spectra of (111) HTO SC using 633 nm laser line in back-scattered geometry showing anomalously scattered modes for $\vec{E}_{in}$ parallel to [1$\bar{1}0$] (top), [11$\bar{2}$] (middle) and in between at 45$^{\circ}$ (bottom).  Modes do not show any clear symmetry as their selection rule appears to be somewhat relaxed. The 0 deg spectra has been fitted using Lorentzian model.  (Analyzer axis $\parallel \vec{E}_{in}$ = 0$^{\circ}$, Analyzer axis $\perp \vec{E}_{in}$ = 90$^{\circ}$)}
\end{figure*}

\clearpage

\begin{figure*}
\centering
\includegraphics[width=0.8\textwidth]{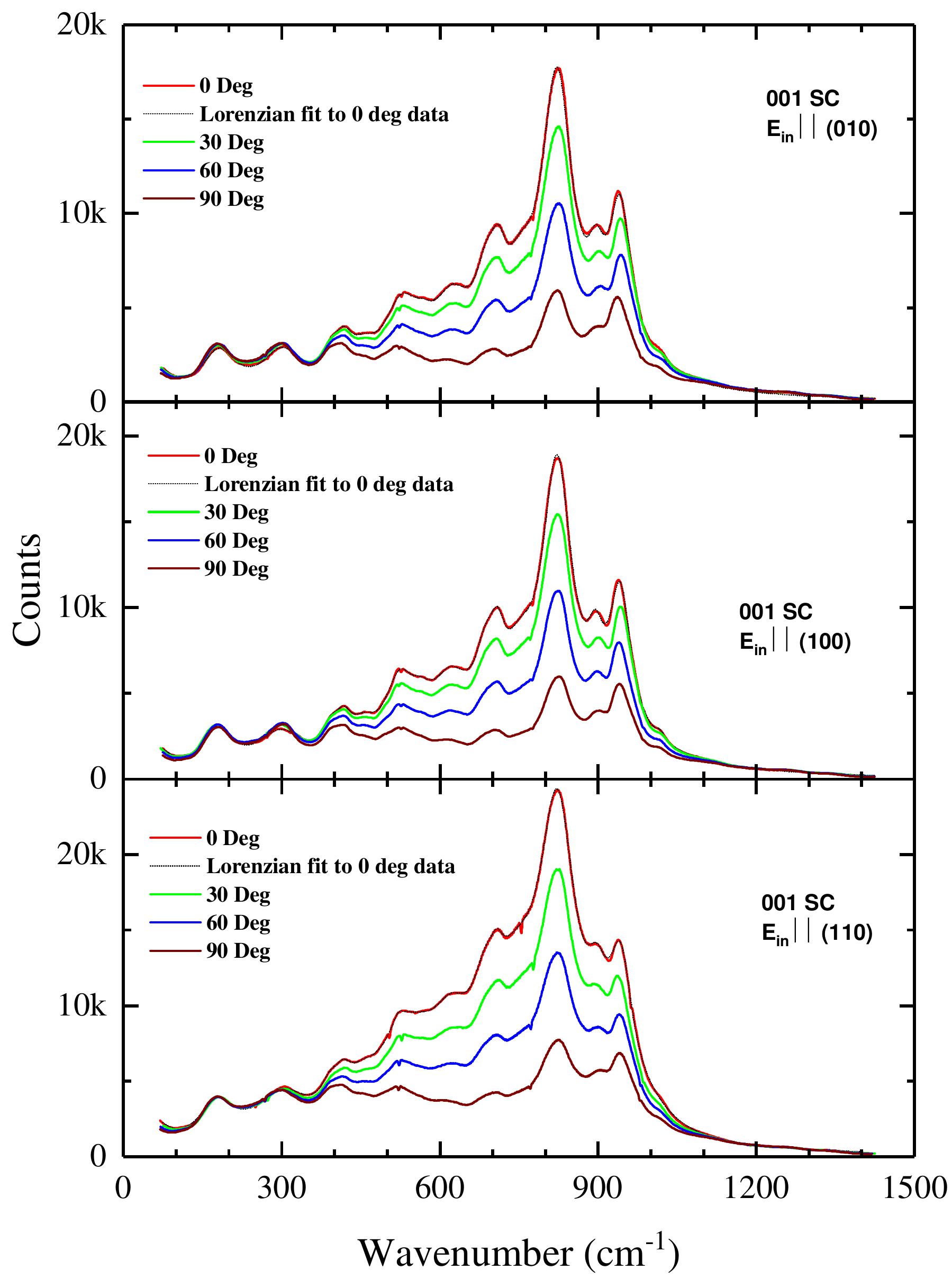}
\caption{\label{001633f}(Color online) 
Polarized resonant Raman spectra of (001) HTO SC using 633 nm laser line in back-scattered geometry showing anomalously scattered modes for $\vec{E}_{in}$ parallel to [010] (top), [100] (middle) and [110] (bottom).   Modes do not show any clear symmetry as their selection rule appears to be somewhat relaxed. The 0 deg spectra has been fitted using Lorentzian model (Analyzer axis $\parallel \vec{E}_{in}$ = 0$^{\circ}$, Analyzer axis $\perp \vec{E}_{in}$ = 90$^{\circ}$)}
\end{figure*}

\clearpage

\begin{figure*}
\centering
\includegraphics[width=0.8\textwidth]{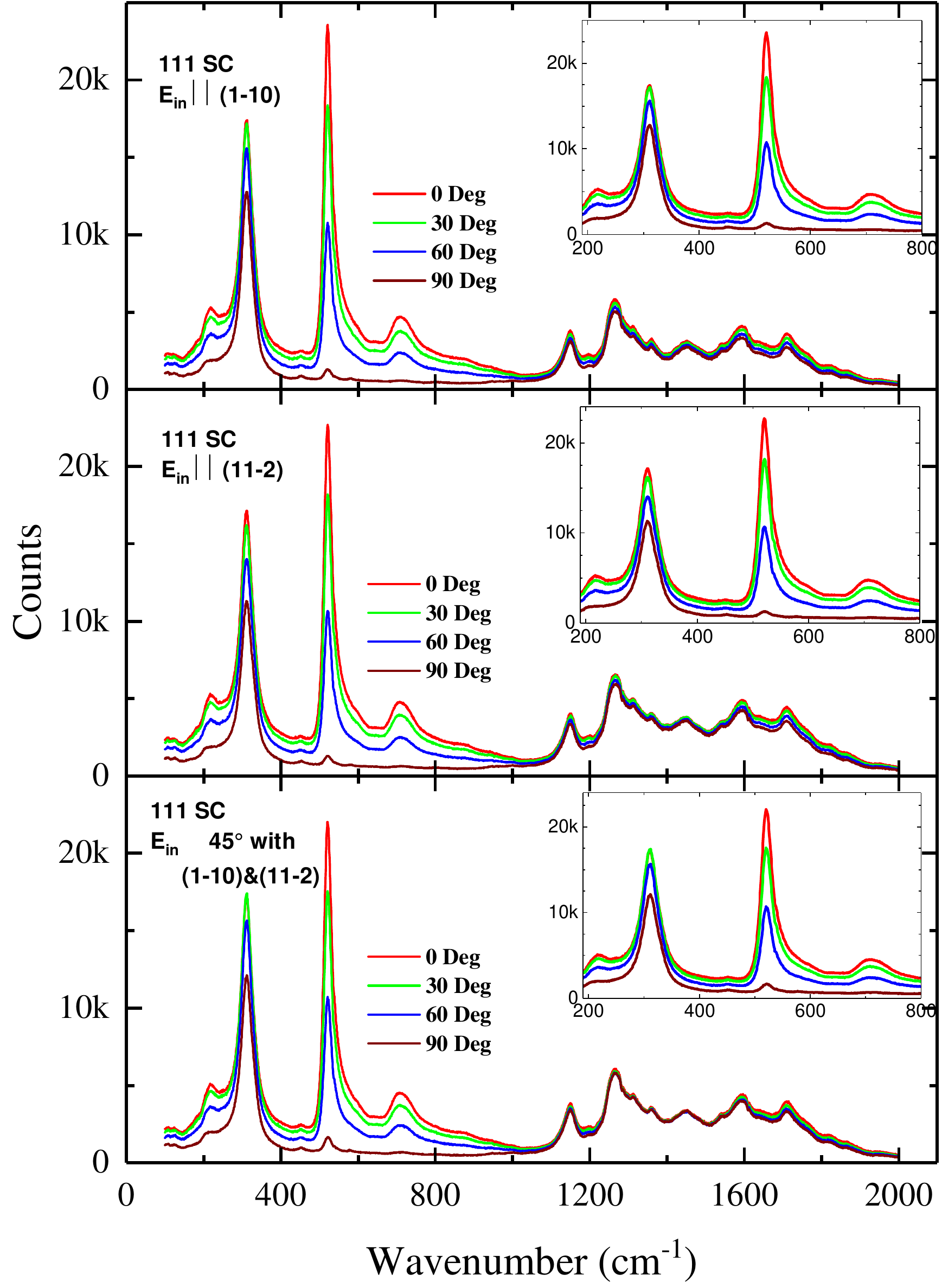}
\caption{\label{111458}(Color online) 
Polarized resonant Raman spectra of (111) HTO SC using 458 nm laser line in back-scattered geometry showing first order fundamental modes and high frequency anomalously scattered modes for $\vec{E}_{in}$ parallel to [1$\bar{1}0$] (top), [11$\bar{2}$] (middle) and in between at 45$^{\circ}$ (bottom). Inset graphs show the zoomed-in spectra at lower frequencies showing first order Raman modes  (Analyzer axis $\parallel \vec{E}_{in}$ = 0$^{\circ}$, Analyzer axis $\perp \vec{E}_{in}$ = 90$^{\circ}$)}
\end{figure*}

\clearpage

\begin{figure*}
\centering
\includegraphics[width=0.8\textwidth]{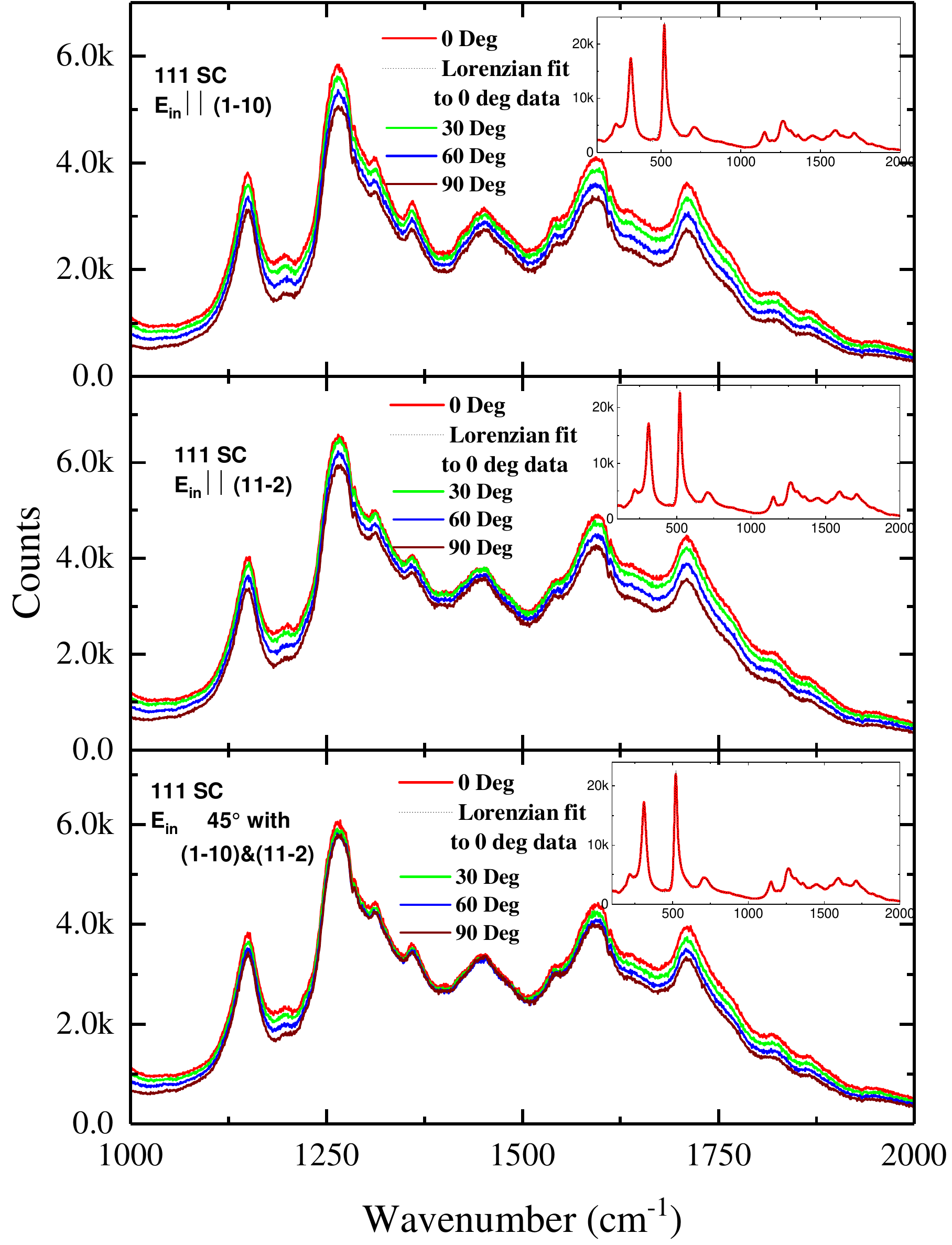}
\caption{\label{111458f}(Color online) 
Polarized resonant Raman spectra of (111) HTO SC using 458 nm laser line in back-scattered geometry showing only the high frequency anomalously scattered modes for $\vec{E}_{in}$ parallel to [1$\bar{1}0$] (top), [11$\bar{2}$] (middle) and in between at 45$^{\circ}$ (bottom). Modes do not show any clear symmetry as their selection rule appears very relaxed. Inset graphs show the 0 deg spectra fitting using Lorentzian model.  (Analyzer axis $\parallel \vec{E}_{in}$ = 0$^{\circ}$, Analyzer axis $\perp \vec{E}_{in}$ = 90$^{\circ}$)}
\end{figure*}

\clearpage

\begin{figure*}
\centering
\includegraphics[width=0.8\textwidth]{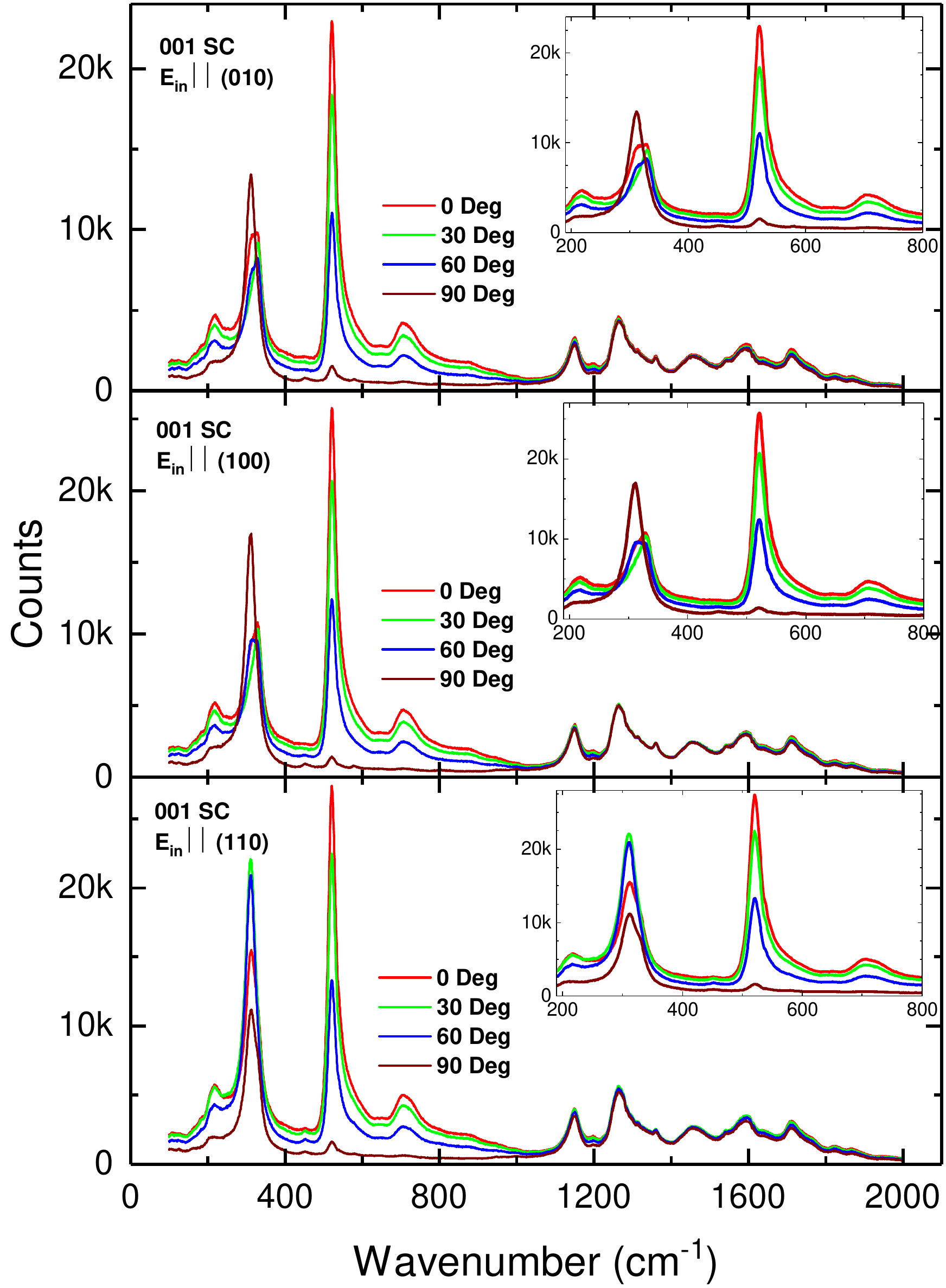}
\caption{\label{001458}(Color online) 
Polarized resonant Raman spectra of (001) HTO SC using 458 nm laser line in back-scattered geometry showing first order fundamental modes and high frequency anomalously scattered modes for $\vec{E}_{in}$ parallel to [010] (top), [100] (middle) and [110] (bottom). Inset graphs show the zoomed-in spectra at lower frequencies showing first order Raman modes  (Analyzer axis $\parallel \vec{E}_{in}$ = 0$^{\circ}$, Analyzer axis $\perp \vec{E}_{in}$ = 90$^{\circ}$)}
\end{figure*}

\clearpage

\begin{figure*}
\centering
\includegraphics[width=0.8\textwidth]{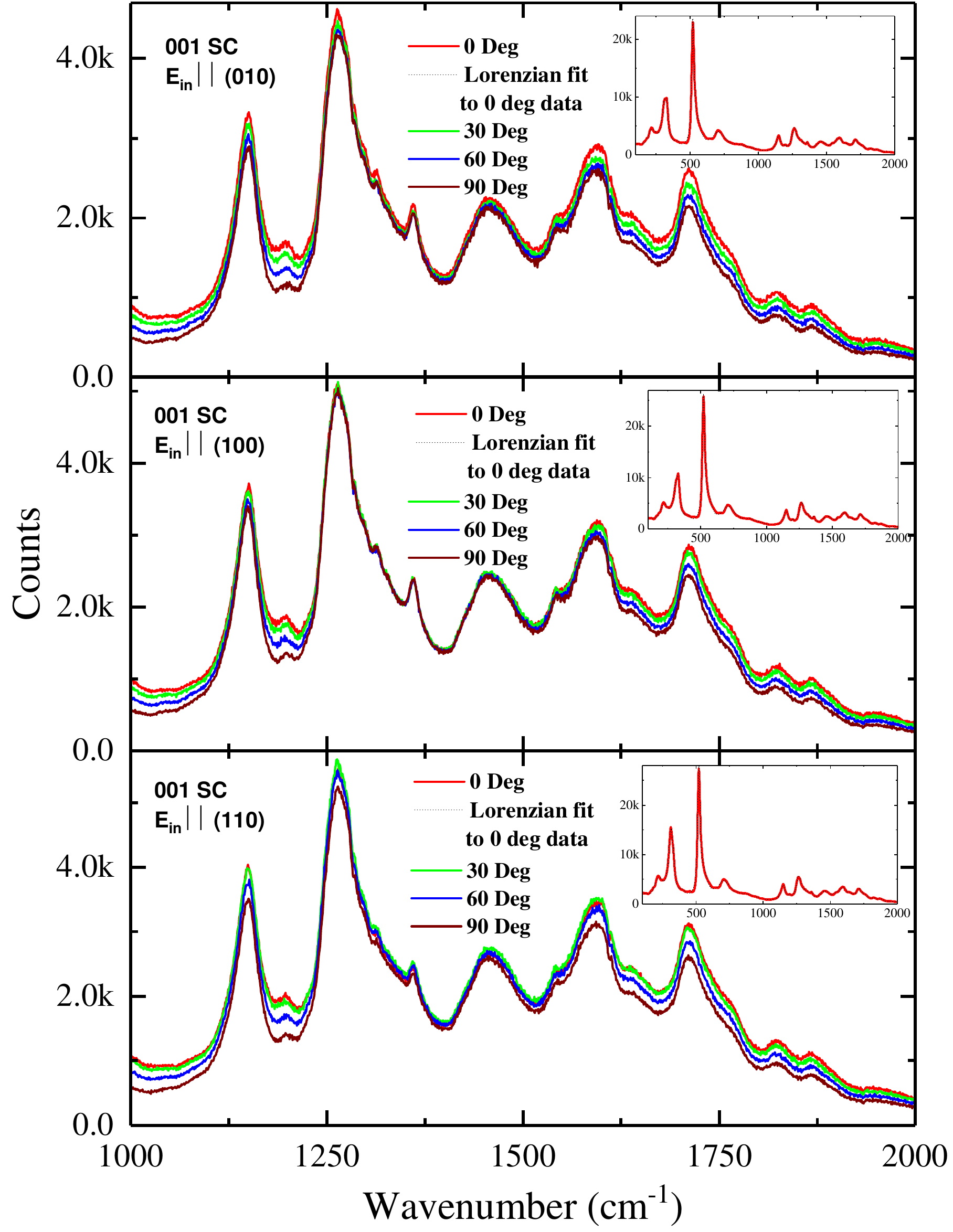}
\caption{\label{001458f}(Color online) 
Polarized resonant Raman spectra of (001) HTO SC using 458 nm laser line in back-scattered geometry showing only the high frequency anomalously scattered modes for $\vec{E}_{in}$ parallel to [010] (top), [100] (middle) and [110] (bottom). Modes do not show any clear symmetry as their selection rule appears very relaxed.  Inset graphs show the 0 deg spectra fitting using Lorentzian model. (Analyzer axis $\parallel \vec{E}_{in}$ = 0$^{\circ}$, Analyzer axis $\perp \vec{E}_{in}$ = 90$^{\circ}$)}
\end{figure*}

\clearpage

\begin{figure*}
\centering
\includegraphics[width=\textwidth]{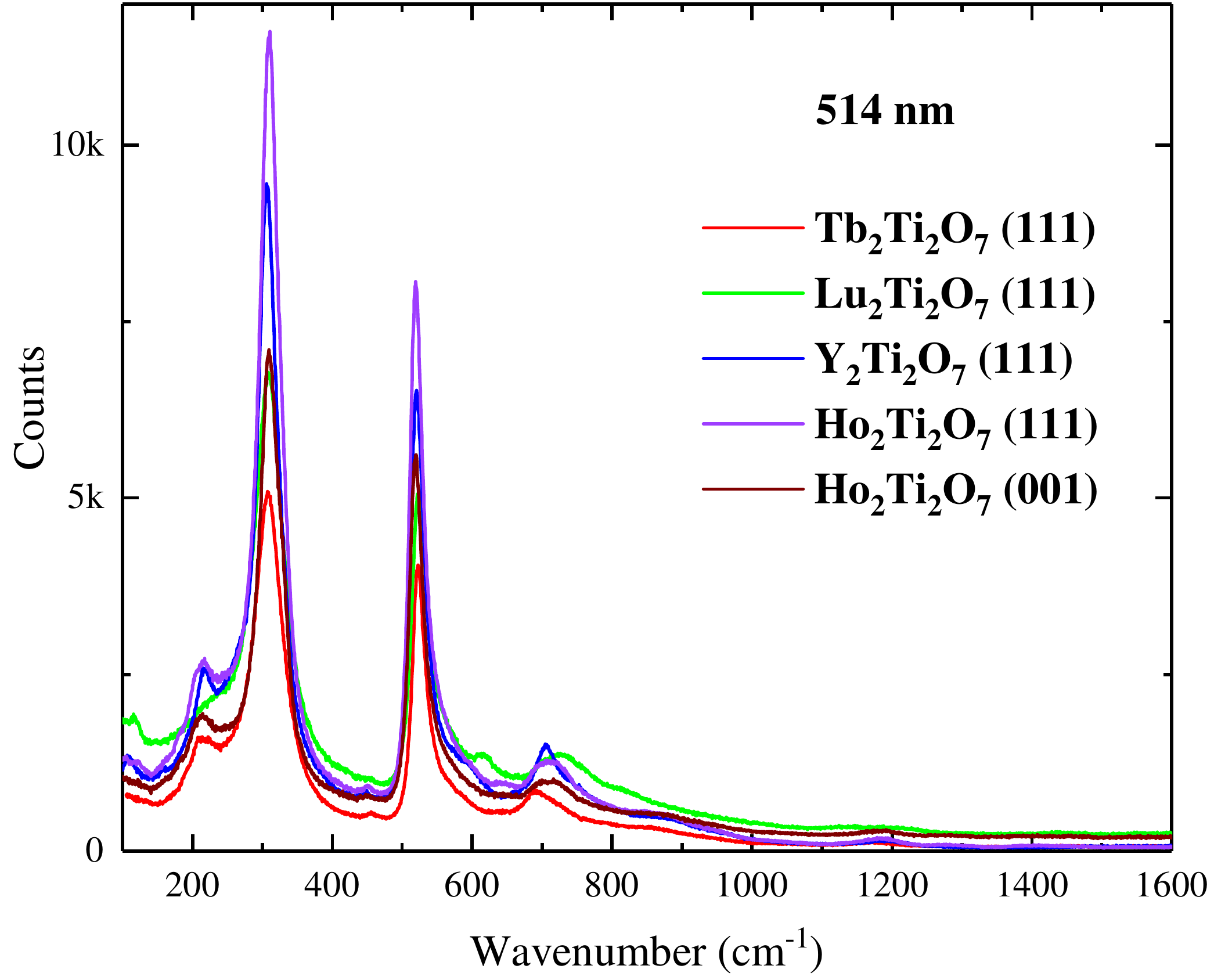}
\caption{\label{514nonpol}(Color online) Unpolarized Raman measurements on several rare-earth pyrochlore single crystals using 514 nm laser excitation. Non-resonant Raman spectra for Ho$_2$Ti$_2$O$_7$ show no difference as compared to any other pyrochlores.}
\end{figure*}
Non-resonant Raman scattering is very similar for all pyrochlores with 514 nm line, all showing characteristic phonon modes as predicted by factor group analysis. Several experimental and theoretical studies on rare-earth pyrochlores has shown similar results.\cite{Lummen,Mkaczka,KUMAR,Ruminy,Kushwaha,Bi,Saha,Sanjuan,BAE,Lee} However when resonance is met using 633 nm and 458 nm laser lines for Ho$_2$Ti$_2$O$_7$, strikingly different Raman cross-section is observed as explained in main text.

\clearpage

\begin{figure*}
\centering
\includegraphics[width=\textwidth]{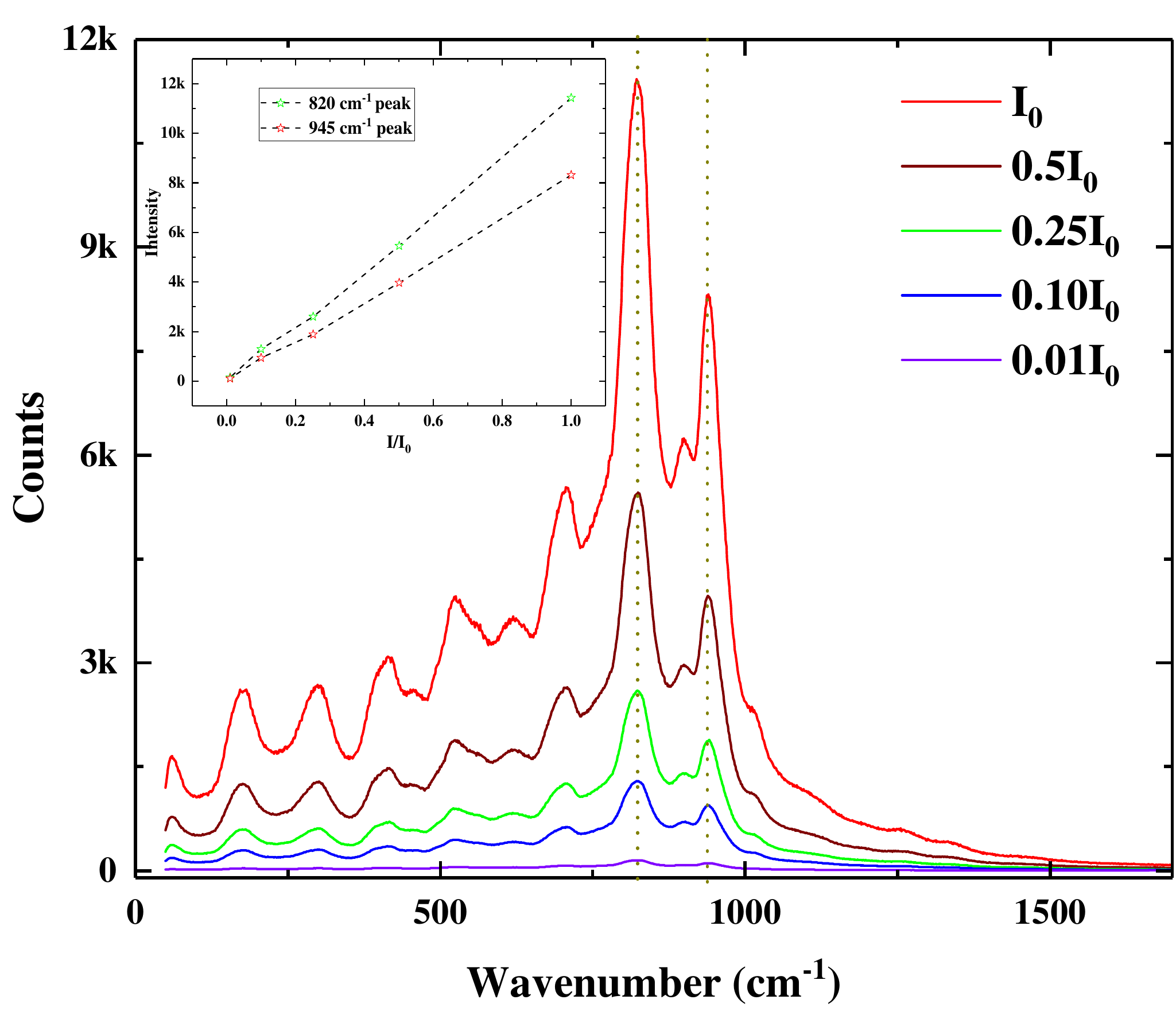}
\caption{\label{Intensity}(Color online) Unpolarized Raman measurements on Ho$_2$Ti$_2$O$_7$ (111) single crystal using 633 nm laser excitation at several filter attenuation settings. Inset graph shows linear trend in output signal at two arbitrarily chosen peaks.}
\end{figure*}
Resonant Raman scattering measurements on Ho$_2$Ti$_2$O$_7$ (111) SC is performed using 633 nm line at several filter-attenuation settings. Output signal at two major peaks has been analyzed and shown in the inset graph. A linear trend between varying input intensity and corresponding output intensity suggests the absence of any non linear processes such as multiphoton absorption, stimulated Raman, thermal effects on sample itself or non-linear optical effects from the optical components in the set-up.

\clearpage

\begin{figure*}
\centering
\includegraphics[width=\textwidth]{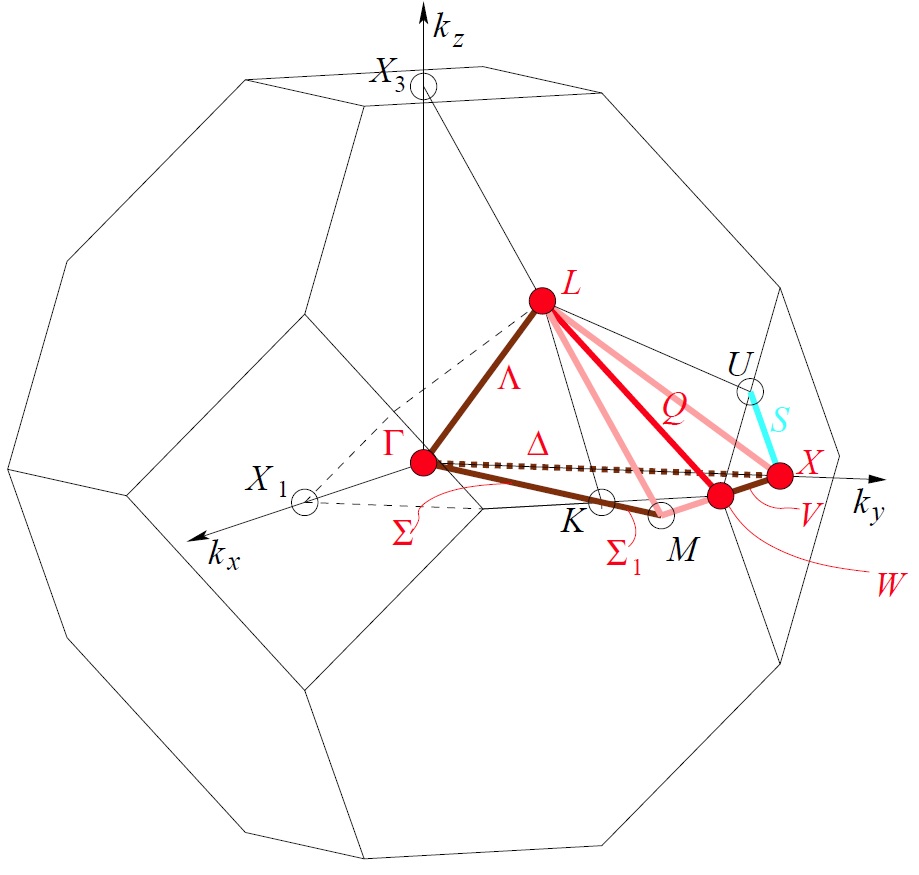}
\caption{\label{BZ}(Color online) Brillouin zone of Ho$_2$Ti$_2$O$_7$ showing all important high symmetry points within the truncated octahedron boundaries. Below is the list of all vibrational modes shown with their associated symmetry and degeneracy as superscript.}
\end{figure*}

\begin{subequations}
\begin{equation}
\Gamma_{3N}= 1A_{1g}^{1}+2E_{g}^{2}+12T_{2g}^{3}+ 24T_{1u}^{3}+ 6T_{1g}^{3}+3A_{2u}^{1}+6E_{u}^{2}+12T_{2u}^{3} 
\end{equation}
\begin{equation}
 \textit{L}= 32L_{1+}^{4}+12L_{2+}^{4}+88L_{3+}^{8}+12L_{1-}^{4}+32L_{2-}^{4}+88L_{3-}^{8}
\end{equation}
\begin{equation}
 \textit{W}= 192W_{1}^{12}+204W_{2}^{12}
\end{equation}
\begin{equation}
\textit{X}= 60X_{1}^{6}+30X_{2}^{6}+48X_{3}^{6}+60X_{4}^{6}
\end{equation}
\end{subequations}

\clearpage
\end{document}